\documentclass[1p, authoryear]{elsarticle}

\usepackage[utf8]{inputenc}

\usepackage{subcaption}
\usepackage{color}

\usepackage{mathtools}
\usepackage{amsmath, amsfonts, amsthm}
\usepackage{bm}

\graphicspath{{img/pdf/}}
\DeclareGraphicsExtensions{.pdf}

\usepackage{algorithm}
\usepackage{algorithmic}
\usepackage[normalem]{ulem}

\usepackage{utilities}
\usepackage{enumitem,booktabs}

\usepackage{tabularx,multirow}

\begin{document}

\title{A review of Gaussian Markov models for conditional independence}

\author{Irene Córdoba\corref{cor}}\ead{irene.cordoba@upm.es}
\author{Concha Bielza\corref{}}\ead{mcbielza@fi.upm.es}
\author{Pedro Larrañaga\corref{}}\ead{pedro.larranaga@fi.upm.es}
\address{Universidad Politécnica de Madrid, Madrid, Spain}

\cortext[cor]{Corresponding author}

\begin{abstract}
	Markov models lie at the interface between statistical independence in a
	probability distribution and graph separation properties. We review model
	selection and estimation in directed and undirected Markov models with
	Gaussian parametrization, emphasizing the main similarities and differences.
	These two model {classes} are similar but not equivalent,
	although they share a common intersection. We present the existing results
	from a historical perspective, taking into account the amount of literature
	existing from both the artificial intelligence and statistics research
	communities, where these models were originated. {We cover
	classical topics such as maximum likelihood estimation and model selection
	via hypothesis testing, but also more modern approaches like regularization
	and Bayesian methods.} We also discuss how { the Markov models
	reviewed fit in the rich hierarchy of other, higher level Markov model
	classes.} Finally, we close the paper overviewing relaxations of the
	Gaussian assumption and pointing out the main areas of application where
	these Markov models are nowadays used.
\end{abstract}

\maketitle

\section{Introduction}
Markov models, or probabilistic graphical models, explicitly establish a
correspondence between statistical independence in a probability distribution
and certain separation criteria holding in a graph. They were originated at the
interface between statistics, where Markov random fields were predominant
\citep{darroch1980}, and artificial intelligence, with a focus on Bayesian
networks \citep{pearl1985b,pearl1986}. These two model { classes} are now considered the
traditional ones, but still are widely applied and nowadays there is a
significant amount of research devoted to them \citep{daly2011,uhler2012}. They
both share the modelling of conditional independences: Bayesian networks relate
them with acyclic directed graphs, whereas in Markov fields they are associated
with undirected graphs. However, the models they represent are only equivalent
under additional assumptions on the respective graphs. 

In this paper, we review the existing methods for model selection and estimation
in undirected and acyclic directed Markov models with a Gaussian
parametrization. The multivariate Gaussian distribution is among the most widely
developed and applied statistical family in this context
\citep{werhli2006,ibanez2015}, and allows for an explicit parametric comparison
of their similarities and differences. The highly interdisciplinary nature of
these Markov model { classes} {has led to a wide range of terminology in methodological
developments and theoretical results}. They have usually been studied
separately, with some exceptions \citep{wermuth1980,pearl1988}, and most
unifying works \citep{sadeghi2014,wermuth2015} are characterized by a high-level
view, where the models are embedded in other, more
expressive { classes}, and the focus is on the properties of these
container { classes}. By contrast, in this paper we {review} them
{from a low-level perspective. In doing so, we use a unified notation that
allows for a direct comparison between the two types of { classes}.
\blue{Furthermore,} throughout each section we explicitly compare {
them}, in terms of both methodological and theoretical developments.} 

{The paper is structured as follows. A historical introduction to Markov models
is presented in Section \ref{sec:his}, emphasizing the different research areas
that contributed to their birth.} Afterwards, preliminary concepts from graph
theory are presented in Section \ref{sec:prel}. In Section \ref{sec:markov},
undirected and acyclic directed Markov model { classes} are introduced, under no
distributional assumptions. This is because many foundational relationships
between them can already be established from this general perspective. Next, we
restrict their parametrization to multivariate Gaussian distributions, and
explore the main derived properties from this in Section \ref{sec:gparam}. We
review maximum likelihood estimation in Section \ref{sec:mle}. These estimates
are used for model selection via hypothesis testing, as we present in Section
\ref{sec:hypo}. When maximum likelihood estimators are not guaranteed to exist,
a popular technique is to employ regularization, which we overview in Section
\ref{sec:reg}. Finally, the alternative Bayesian approach for model selection
and estimation is treated in Section \ref{sec:bayes}. {We explore
the relationship of Gaussian acyclic directed and undirected Markov models with
other, higher level model classes in Section \ref{sec:higher}.} Alternatives to the
Gaussian distribution are discussed in Section \ref{sec:alt}. We close
the paper discussing the main real applications of the Gaussian Markov model
{ classes}
reviewed in Section \ref{sec:app}.

\section{{A historical perspective}}\label{sec:his}
\begin{figure}[h!]
	\centering
	\includegraphics{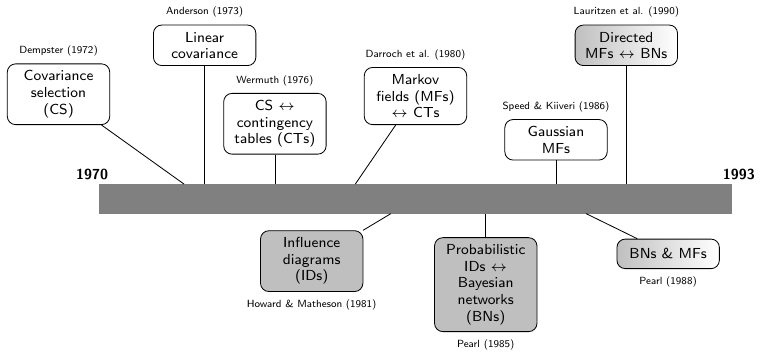}  
	\caption{Timeline on the origins of Gaussian Markov models. Papers from the
	statistical community appear at the top, while papers from other research
	areas appear below. Thematically, gray filled squares are papers about
	acyclic directed Markov models, the white ones are about the undirected
	case, and those gradient filled treat both {classes}. }
	\label{fig:timeline}
\end{figure}

We will now introduce the main terminology for Gaussian Markov models that can
be found nowadays, from a historical perspective. In Figure \ref{fig:timeline}
we have depicted a timeline on the origins of Markov models, containing most of
the key works we will refer to in this section.

Undirected Markov models for conditional independence are the oldest type of
Markov models, preceded only by special cases such as the Ising model for
ferromagnetic materials \citep{kindermann1980,isham1981}. In fact, they are a
generalization of the Ising model, which is at the same time a generalization of
Markov chains. Originally, undirected Markov models were called \textit{Markov
random fields} \citep{grimmett1973}, since they generalized the correspondence
between Gibbs measures \citep{besag1974} and Markov properties. The terminology
\textit{graphical model} was not introduced until \citet{darroch1980} linked the
graphical ideas for contingency tables with Markov properties of discrete Markov
fields. Furthermore, we also find them called \textit{Markov networks}
\citep{pearl1988}, from researchers in artificial intelligence, as a parallel to
the terminology \textit{Bayesian networks}, used for acyclic directed Markov
fields. 

Regarding the Gaussian parametrization, we can find that one of the first
{works} to impose some structure on the covariance matrix of a multivariate
{Gaussian} distribution, in order to reduce the number of parameters to be
estimated, was \citet{anderson1973}. He considered the mean vector and
covariance matrix to be linear combinations of known linearly independent
vectors and matrices, respectively. Closely following this work was
\citet{dempster1972}, who suggested to {estimate} the inverse of the covariance
matrix (concentration matrix) {by assuming certain entries equal to} zero,
motivated by the representation of the multivariate Gaussian distribution as an
exponential family. His work was later referred to as \textit{covariance
selection} models. Interestingly, although Dempster did not have any graphical
interpretation in mind, such zero entries in the concentration matrix are
directly associated with missing edges in an undirected Gaussian Markov models,
and these correspondence was analysed some years later in \citet{wermuth1976}.
This is why, even nowadays, these Markov models with a Gaussian parameterization
are sometimes called covariance selection models.

{Acyclic} digraphs, {in contrast}, were intensely used as models for
multivariate probability distributions after the definition of \textit{influence
diagrams}. These are used to model decision-making processes, and were
introduced by Howard and Matheson in 1981 (article reprinted in
\citet{howard2005}). Their probabilistic reduction coincides with acyclic
directed Markov models, and was subsequently extensively studied by Pearl
\citep{pearl1988}, who renamed probabilistic influence diagrams as
\textit{Bayesian networks} or \textit{influence networks} \citep{pearl1985b}.
Some researchers working on Markov fields also developed theory regarding these
directed counterparts, calling them \textit{directed Markov fields}
\citep{lauritzen1990}.

Earlier works than the previously outlined, employing or referencing acyclic
directed Markov models, are available. \citet{wermuth1980} implicitly studied
them in the Gaussian case as \textit{linear recursive regression systems},
although the main focus was rather on covariance selection models. In fact, we
can trace the use of directed graphs as graphical models for dependencies among
random variables at least to the work of geneticist Sewall Wright in 1918, who
developed the method of \textit{path coefficients} \citep{wright1934}, nowadays
known as \textit{path analysis}. Linearly related variables {were represented
using} a directed {acyclic} graph{, whereas their correlation was represented by
bi-directed edges joining them}.

\section{Graph preliminaries}\label{sec:prel}
\begin{figure}[h!]
	\begin{subfigure}[b]{0.3\textwidth}
		\centering
		\includegraphics{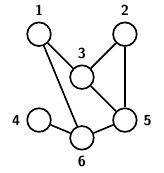}  
		\caption{Undirected graph}
		\label{fig:ug}
	\end{subfigure}
	\begin{subfigure}[b]{0.3\textwidth}
		\centering
		\includegraphics{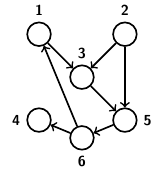}
		\caption{Cyclic digraph}
		\label{fig:dg}
	\end{subfigure}
	\begin{subfigure}[b]{0.3\textwidth}
		\centering
		\includegraphics{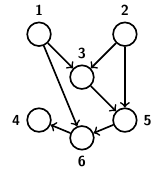}
		\caption{Acyclic digraph}
		\label{fig:dag}
	\end{subfigure}
	\caption{Examples of an undirected graph and two digraphs.}
\end{figure}

A graph is defined as a pair $\gdesc{\graph{G}}{V}{E}$ where $V$ is the vertex set and $E$ is the edge set. Throughout all the paper, and unless otherwise stated, the graphs will be labelled and simple, which means that the elements in $V$ are labelled, for example, as $1, \ldots, p$; and $E$ {is formed by pairs of distinct elements in $V$}. A graph is called \textit{undirected} {if these latter pairs are unordered} ($E \subseteq \set{\set{u, v} \st u, v \in V}$), and \textit{directed} or \textit{digraph} {otherwise} ($E \subseteq \set{(u, v) \st u, v \in V}$). Edges $\set{u, v}$ in an undirected graph are usually denoted as $uv$ {and graphically represented as a line} (see Figure \ref{fig:ug}); while in a digraph they are called \textit{arcs} or \textit{directed edges} {and represented as arrows} (Figure {\ref{fig:dg} and} \ref{fig:dag}). 

\subsection{Undirected graphs}

In an undirected graph $\gdesc{\graph{G}}{V}{E}$, if $uv\in E$, $u$ and $v$ are called \textit{neighbours}. For $v\in V$, the set of its neighbours is denoted as $\nei(v)$, and the \textit{closure} of $v$ is $\cl(v) \sdefeq \set{v}\cup\nei(v)$. $\graph{G}$ is called \textit{complete} if for every $u,v \in V$, $uv \in E$. A maximal $C\subseteq V$ such that $\graph{G}_C$ is complete is called a \textit{clique}. Let $\gdesc{\graph{H}}{\svertex{V}{\graph{H}}}{\sedges{E}{\graph{H}}}$ be another undirected graph. $\graph{H}$ is a \textit{sub-graph} of $\graph{G}$ (written as $\graph{H}\subseteq\graph{G}$) if $\svertex{V}{\graph{H}}\subseteq V$ and $\sedges{E}{\graph{H}}\subseteq E$. If $\sedges{E}{\graph{H}} = \{uv\in E: u, v\in \svertex{V}{\graph{H}}\}$, then $\graph{H}$ is called the \textit{induced sub-graph} and denoted $\graph{G}_{\svertex{V}{\graph{H}}}$. 

A \textit{walk} between $u$ and $v$ is an ordered sequence of vertices $(u =) u_0, u_1,\ldots,u_{k-1}, u_k (= v)$ where $u_{i-1}u_i \in E$ for $i\in \{1, \ldots, k\}$. The number $k$ is called the \textit{length} of the walk. If $u = v$ the walk is \textit{closed}, and when $u_0, \ldots, u_{k - 1}$ are distinct, the walk is called a \textit{path}. A closed path of length $k \ge 3$ is called a \textit{cycle} or $k$-cycle. $\graph{G}$ is called \textit{chordal} or \textit{triangulated} if all minimal $k$-cycles are of length $k=3$. A \textit{chordal cover} of a graph $\graph{G}$ is a graph $\gchord{\graph{G}}$ such that $\graph{G} \subseteq \gchord{\graph{G}}$ and $\gchord{\graph{G}}$ is chordal. 

$S \subseteq V$ \textit{separates} $u$ and $v$ in $\gdesc{\graph{G}}{V}{E}$ if there is no path between $u$ and $v$ in the sub-graph $\graph{G}_{V\setminus S}$. If we consider $A,B,S \subseteq V$, $A$ and $B$ are said to be \textit{separated} by $S$ if $u$ and $v$ are separated by $S$ for all $u\in A$, $v \in B$. Let $V$ be partitioned into disjoint sets $A,B,S\subseteq V$. $(A,B,S)$ is called a \textit{decomposition} of $\graph{G}$ if $S$ separates $A$ and $B$ in $\graph{G}$ and $\graph{G}_S$ is complete. If $A\neq\emptyset$ and $B\neq\emptyset$ the decomposition is said to be \textit{proper}. An undirected graph is \textit{decomposable} if (i) it is complete or (ii) {it} admits a proper decomposition into \textit{decomposable} sub-graphs. An undirected graph is decomposable if and only if it is chordal. 

\subsection{Acyclic digraphs}

In a digraph $\gdesc{\graph{D}}{V}{A}$ the definitions of (induced) sub-graph, walk, path, and cycle are analogous to the undirected case. The undirected graph $\gskel{\graph{D}} \sdefeq (V, A^U)$ with $A^U \sdefeq \set{uv: (u,v)\in A}$ is called the \textit{skeleton} of $\graph{D}$, and $\graph{D}$ is one of its \textit{orientations}. {A digraph $\graph{D}$ is said to be \textit{complete} when $\gskel{\graph{D}}$ is complete.}

In the following, assume that $\graph{D}$ is acyclic {(see Figure \ref{fig:dg} for a cyclic digraph, and Figure \ref{fig:dag} for an acyclic one)}. The \textit{parent set} of $v \in V$ is $\pa(v) \sdefeq \set{u\in V: (u,v)\in A}$; conversely, the \textit{child set} is $\ch(v) \sdefeq \{u\in V: (v,u)\in A\}$. The \textit{ancestors} of $v$, $\an(v)$, are those $u\in V$ such that there exists a directed path from $u$ to $v$; the \textit{descendants} of $v$, $\de(v)$, are those $u\in V$ such that there exists a directed path from $v$ to $u$. We will {let} $\nd(v) \sdefeq V\setminus(\{v\}\cup\de(v))$ {be} the {set of} \textit{non-descendants} of $v$ ${\in V}$, and $\An(A) \sdefeq A \cup (\cup_{a\in A}\an(a))$ {the \textit{ancestral set} of} $A \subseteq V$. Note that a total order $\prec$ can be defined over the set of vertices $V$ in an acyclic digraph $\gdesc{\graph{D}}{V}{A}$, such that if $(u, v) \in A$, then $u \prec v$. This ordering is usually called \textit{ancestral}, and it is a linear extension of the partial order naturally defined as $u \preceq v$ if $u \in \an(v)$. For $v \in V$, the set of \textit{successors} of $v$ {with respect to $\prec$} is $\su(v) = \set{u\in V: u \succ v}$; the set of \textit{predecessors} of $v$ is $\pr(v) = \set{u\in V: u \prec v}$.

Finally, let $u,{w_1,w_2} \in V$ with ${(w_1,u), (w_2,u)} \in A$ {and $(w_1, w_2), (w_2, w_1) \notin A$} (see vertices $1$, $2$ and $3$ in Figure \ref{fig:dag}). Such configurations are usually called \textit{v-structures} and {denoted as} {$\vstruc{w_1}{w_2}{u}$}. The \textit{moral graph} of $\graph{D}$ is defined as the undirected graph $\gmoral{\graph{D}} = (V, {A^m})$ with $A^m \sdefeq A^U\cup \set{w_1w_2:{\vstruc{w_1}{w_2}{u}}\text{ for some }u\in V}$.

\section{Undirected and acyclic directed Markov model { classes}}\label{sec:markov}
The Markov model {classes} we will review associate conditional independences in random
vectors $\vrand = (X_1, \ldots, X_p)^t$ with undirected graph and acyclic
digraph separation properties. This is made explicit via the \textit{Markov
properties} of the distribution of $\vrand$, which are in turn based on what are
known as \textit{independence relations}. 

In the following, for arbitrary $I\subseteq\set{1,\ldots,p}$, we will denote the
$\card{I}$-dimensional sub-vector of $\vrand$ as $\vrand_{I} \sdefeq (X_i)_{i\in
I}$. Conditional independence will be expressed as in \citet{dawid1979}:
$\ci{\vrand_I}{\vrand_J}{\vrand_K}$ represents the statement `$\vrand_I$ is
conditionally independent from $\vrand_J$ given $\vrand_K$' \citep[see,
e.g.][\S1.3]{studeny2018}. 

\subsection{Independence relations}\label{sec:prel:dep}

A{n} \textit{independence relation} {over} a set $V = \set{1,\ldots,p}$ {is} a
collection $\sirel{I}$ of triples $(A, B, C)$ where $A$, $B$ and $C$ are
pairwise disjoint subsets of $V$. It is called a \textit{semi-graphoid} when the
following conditions are met,
\begin{align*}
&\text{ if }\itriple{I}{A}{B}{C}{\text{ then }} \itriple{I}{B}{A}{C},\\
&\text{ if }\itriple{I}{A}{B \cup C}{D} {\text{ then }} \itriple{I}{A}{C}{D} \text{ and } \itriple{I}{A}{B}{C \cup D},\\
&\text{ if }\itriple{I}{A}{B}{C \cup D} \text{ and } \itriple{I}{A}{C}{D} {\text{ then }} \itriple{I}{A}{B \cup C}{D};
\end{align*}
and a \textit{graphoid} when it additionally satisfies that if
$\itriple{I}{A}{B}{C \cup D}$ and $\itriple{I}{A}{C}{B \cup D}$ then
$\itriple{I}{A}{B \cup C}{D}$ \citep{pearl1987}.

Independence relations arise in different contexts relevant for Markov models.
Specifically, an independence relation $\sirel{I}$ over $V = \set{1, \ldots, p}$
is said to be \textit{induced} by
\begin{itemize}
	\item an undirected graph $\gdesc{\graph{G}}{V}{E}$ if $\itriple{I}{A}{B}{S}$ $\iff$ $A$ and $B$ are separated by $S$ in $\graph{G}$,
	\item an acyclic digraph $\gdesc{\graph{D}}{V}{A}$ if $\itriple{I}{A}{B}{S}$ $\iff$ $A$ and $B$ are separated by $S$ in $\gmoral{(\graph{D}_{\An(A \cup B \cup S)})}$,
	\item a $p$-dimensional random vector $\vrand$ if $\itriple{I}{A}{B}{S} \iff \ci{\vrand_A}{\vrand_B}{\vrand_S}$.
\end{itemize}
Graph-induced independence relations are always graphoids, while probabilistic
ones are always semi-graphoids and require additional {assumptions} on the
probability spaces involved to be graphoids \citep{dawid1980}.
{See \citet{studeny2018} \S1.5 and \S1.11 for a detailed exposition
of graphoid theory and how to compute and represent their closures, that is, all
the triplets that can be derived from a given independence relation by using the
graphoid axioms.}

The core of Markov model { classes} is the relationship between induced independence
relations, which we will denote as $\firel{I}{\cdot}$ with the argument being
the inducing element. Specifically, if $\graph{G}$ is an undirected (acyclic
directed) graph, an undirected (directed) \textit{Markov model} is defined as 
\[
	\gmm{M}{\graph{G}} \sdefeq \set{\pdist{\vrand} \st \firel{I}{\graph{G}} \subseteq \firel{I}{\vrand}}.
\]
where the random vectors $\vrand$ are defined over the same
probability space and $\pdist{\vrand}$ denotes their
distribution. These { classes} are non-empty
\citep{geiger1990b,geiger1993}; that is, for any undirected or
acyclic directed graph, we can always find a probability
distribution whose independence model contains the one generated by the graph.

{ Graphoids can be generalised to what are known as
\textit{separoids} \citep{dawid2001}, which are algebraic structures usually
appearing whenever a notion of `irrelevance' is being mathematically treated
\citep[see, e.g.][\S1.1.3]{studeny2018}. Further research on these axiom systems
from an abstract point of view could shed more light {on} how the apparently
different mathematical contexts {in which} such structures arise are related,
and also provide an explicit bridge between them and the recently defined
\textit{independence logic} \citep{graedel2013}, closely related.}

\subsection{Markov properties}\label{sec:mp}
When a distribution $\pdist{\vrand}$ belongs to $\gmm{M}{\graph{G}}$ for an
undirected or acyclic directed graph $\graph{G}$, it is said that
$\pdist{\vrand}$ is \textit{globally $\graph{G}$-Markov} or satisfies the
\textit{global Markov property} with respect to $\graph{G}$. There are other
weaker Markov properties that usually allow to simplify the model. Specifically,
if $\gdesc{\graph{G}}{V}{E}$ is an undirected graph, then the probability
distribution $\pdist{\vrand}$ of $\vrand$ is said to be
\begin{itemize}
	\item \textit{pairwise $\graph{G}$-Markov} if $\ci{X_u}{X_v}{\vrand_{V\setminus\set{u,v}}}$ for all $uv\notin E$,
	\item \textit{locally $\graph{G}$-Markov} if $\ci{X_v}{\vrand_{V\setminus \cl(v)}}{\vrand_{\nei(v)}}$ for all $v \in V$;
\end{itemize}
whereas if $\graph{G}$ is an acyclic digraph, then $\pdist{\vrand}$ is called
\begin{itemize}
	\item \textit{pairwise $\graph{G}$-Markov} if $\ci{X_u}{X_v}{\vrand_{\nd(u)\setminus\set{v}}}$ for all $u\in V, v\in\nd(u)\setminus\pa(u)$;\label{dag:pairwise}
	\item \textit{locally $\graph{G}$-Markov} if $\ci{X_v}{\vrand_{\nd(v)\setminus\pa(v)}}{\vrand_{\pa(v)}}$ for all $v \in V$.\label{dag:local}
\end{itemize}
The three Markov properties are equivalent when $\graph{G}$ is acyclic directed
\citep{lauritzen1990}, while if $\graph{G}$ is undirected this equivalence is
only guaranteed when $\firel{I}{\vrand}$ is a graphoid \citep{pearl1988}. A
sufficient condition for this to happen is that $\pdist{\vrand}$ admits a
continuous and strictly positive density. This result was proved in different
forms by several authors, but {it} is usually attributed to
\citet{hammersley1971}, {who} were the first to outline the proof for the
discrete case \citep{speed1979}. It relies on an additional characterization of
a probability distribution with respect to $\graph{G}$: denoting as
$\scliques{\graph{G}}$ the class of cliques of $\graph{G}$, the density function
$f$ of $\pdist{\vrand}$ is said to \textit{factorize according to $\graph{G}$}
when there exists a set $\{\psi_C(\vecb{x}_C): C\in\scliques{\graph{G}},\psi_C
\geq 0\}$ such that
\begin{equation}\label{eq:ug:fact}
	f(\vecb{x}) = \prod_{C\in\scliques{\graph{G}}}\psi_C(\vecb{x}_C).
\end{equation}
When \eqref{eq:ug:fact} holds, then $\pdist{\vrand}$ is globally
$\graph{G}$-Markov, while if $f$ is continuous and strictly positive, the
pairwise Markov property implies \eqref{eq:ug:fact}, which gives the equivalence
of Markov properties. 
{ Positivity is a straightforward sufficient condition for checking
whether an independence model originated from a distribution is a graphoid.
Necessary and sufficient conditions are given in measure theoretic terms by
\citet{dawid1980}, and recently by \citet{peters2014} in terms of special
functions over the sample space.}

Finally, recall that the nodes of an acyclic digraph $\gdesc{\graph{D}}{V}{A}$
can be totally ordered such that if $(u, v) \in A$, then $u \in \pr(v)$. This
gives rise to another Markov property, exclusive for acyclic digraphs:
$\pdist{\vrand}$ is said to be \textit{ordered $\graph{D}$-Markov} if
$\ci{X_v}{\vrand_{\pr(v)\setminus\pa(v)}}{\vrand_{\pa(v)}}$ for all $v\in V$.
This property is also equivalent to the global, local and pairwise Markov
properties \citep{lauritzen1990}. {The classical theory of
undirected and acyclic directed Markov properties can be found in
\citet{lauritzen1996}, whereas \citet{studeny2018} \S1.7 and \S1.8 provides a
recent overview.}

\subsection{Independence and Markov equivalence}
When the Markov models defined by two graphs $\graph{G}$ and
$\graph{\tilde{G}}$, with the same vertex set $V$, coincide,
such graphs are said to be \textit{Markov equivalent}. A
simpler notion, which implies Markov equivalence, is
\textit{independence equivalence}, holding when
$\firel{I}{\graph{G}} = \firel{I}{\graph{\tilde{G}}}$.
Independence equivalence is implied by Markov equivalence
under fairly general circumstances
\citep[][\S6.1]{studeny2005}, which is why most authors treat
them as the same notion. These equivalences allow to choose
the most suited graph for the Markov model. 

We will first characterize equivalence within undirected
graphs. For each graphoid $\sirel{I}$ over $V$ there exists a
unique edge-minimal undirected graph $\graph{G}$ such that
$\firel{I}{\graph{G}} \subseteq \sirel{I}$ \citep{pearl1987}.
It follows that $\firel{I}{\graph{G}} =
\firel{I}{\tilde{\graph{G}}}$ (independence equivalence) if
and only if $\graph{G}$ and $\tilde{\graph{G}}$ are identical.
Furthermore, if we assume that $\firel{I}{\vrand}$ is a
graphoid for all $\pdist{\vrand} \in \gmm{M}{\graph{G}}$, then
a unique edge-minimal $\graph{\tilde{G}}$ exists, with
$\graph{\tilde{G}} \subseteq \graph{G}$, such that
$\gmm{M}{\graph{G}} = \gmm{M}{\graph{\tilde{G}}}$ (Markov
equivalence); that is, a unique undirected graph can be chosen
as representative of each undirected Markov model.

In contrast, acyclic digraphs are not, in general, unique representations of a
Markov model, since $\firel{I}{\graph{D}} = \firel{I}{\tilde{\graph{D}}}$ if and
only if $\graph{D}$ and $\tilde{\graph{D}}$ have the same skeleton and the same
v-structures \citep{verma1991}. However, unique representatives can be
constructed: let $\sacdig{p}$ be the set of acyclic digraphs over $V = \set{1,
\ldots, p}$ and define an equivalence relation $\srel$ in $\sacdig{p}$ as
$\graph{D} \srel \graph{\tilde{D}} \iff \firel{I}{\graph{D}} =
\firel{I}{\graph{\tilde{D}}}$. The quotient space of $\srel$ is
$\sqrel{\sacdig{p}} =  \{\eqclass{\graph{D}} \st \graph{D} \in \sacdig{p}\}$,
where $\eqclass{\graph{D}} \sdefeq \{\graph{\tilde{D}} \in \sacdig{p}  \st
\graph{\tilde{D}} \srel \graph{D}\}$ is the \textit{Markov equivalence class};
indeed, $\gmm{M}{\graph{\tilde{D}}} = \gmm{M}{\graph{D}}$ for all
$\graph{\tilde{D}} \in \eqclass{\graph{D}}$, that is, $\eqclass{\graph{D}}$ is
the unique representative of the directed Markov model. 

{ 
The asymptotic ratio $l =
\lim_{p\tendsto\infty}\card{\sacdig{p}}/\card{\sqrel{\sacdig{p}}}$ influences
the computational gain obtained by using $\sqrel{\sacdig{p}}$ instead of
$\sacdig{p}$ as a search space for model selection. {\citet{steinsky2004}
analytically calculates an upper bound of $l$ as $13.65$}. Exact computations
{by} \citet{gillispie2002}, for $p \leq 10$, and approximations {by}
\citet{sonntag2015}, up to $p = 31$, seem to indicate that $l \sim 3.7${.
H}owever, its analytical deduction remains an open problem. Note that the
computational gain is not only influenced by $l$, but also by other factors,
such as how the element size in $\sqrel{\sacdig{p}}$ is distributed. An
algorithm to compute such sizes can be found in \citet{he2016b}. Recently,
\citet{radhakrishnan2018} have provided tight lower and upper bounds on the number and
size of Markov equivalence classes when $\graph{D}$ is a tree.}

Finally, we will characterize equivalence between directed and undirected
graphs, firstly obtained by \citet{wermuth1980} for multivariate Gaussian
distributions, \citet{wermuth1983} for contingency tables, and generalized in
\citet{frydenberg1990} for graphoid-inducing distributions. When $\graph{G}$ is
an undirected graph, $\gmm{M}{\graph{G}} = \gmm{M}{\graph{D}}$ for some acyclic
digraph $\graph{D}$ if and only if $\graph{G}$ is chordal. Conversely, an
acyclic digraph $\graph{D}$ is Markov equivalent to its skeleton
$\gskel{\graph{D}}$ if and only if $\graph{D}$ contains no v-structures.
Furthermore, a relation with the moral graph can be established, which requires
an analogous to \eqref{eq:ug:fact}: a density function $f$ is said to
\textit{recursively factorize} according to $\graph{D}$ when
\begin{equation*}
	f(\vecb{x}) = \prod_{v\in V}f(x_v \sgiven \vecb{x}_{\pa(v)}).
\end{equation*}
This characterization is equivalent to the Markov properties, and also implies
that $f$ factorizes as in \eqref{eq:ug:fact} with respect to the moral graph
$\gmoral{\graph{D}}$ \citep{lauritzen1990}. This means that $\pdist{\vrand}$ is
always globally $\gmoral{\graph{D}}$-Markov for continuous $\vrand$, and thus
$\gmm{M}{\graph{D}} \subseteq \gmm{M}{\gmoral{\graph{D}}}$, with the equality
only holding when $\gmoral{\graph{D}} = \gskel{\graph{D}}$.

\begin{figure}[h!]
	\begin{subfigure}[b]{0.18\textwidth}
		\centering
		\includegraphics{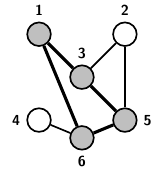}
		\caption{Chordless cycle}
		\label{fig:ug:chordless}
	\end{subfigure}
	\begin{subfigure}[b]{0.18\textwidth}
		\centering
		\includegraphics{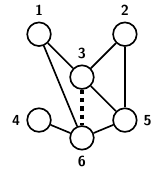}
		\caption{Chordal cover}
		\label{fig:ug:chordalcover}
	\end{subfigure}
	\begin{subfigure}[b]{0.18\textwidth}
		\centering
		\includegraphics{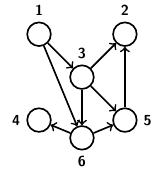}
		\caption{Orientation}
		\label{fig:ug:dageq}
	\end{subfigure}
	\begin{subfigure}[b]{0.18\textwidth}
		\centering
		\includegraphics{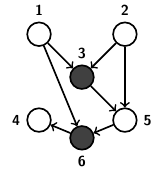}
		\caption{$V$-structures}
		\label{fig:dag:v-struct}
	\end{subfigure}
	\begin{subfigure}[b]{0.18\textwidth}
		\centering
		\includegraphics{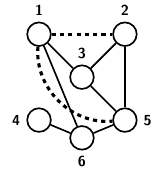}
		\caption{Moral graph}
		\label{fig:dag:moral}
	\end{subfigure}
	\caption{Markov equivalence.}
	\label{fig:dag:eq}
\end{figure}

\begin{exm}
	An illustration of the previous concepts can be found in Figure
	\ref{fig:dag:eq}. The graph in \ref{fig:ug:chordless} is not chordal, and
	thus there is no Markov equivalent acyclic digraph.
	\ref{fig:ug:chordalcover} is a chordal cover of \ref{fig:ug:chordless}, and
	a Markov equivalent orientation is depicted in \ref{fig:ug:dageq}. The
	acyclic digraph in \ref{fig:dag:v-struct} has v-structures, emphasized in
	dark grey, and thus cannot be Markov equivalent to its skeleton
	(\ref{fig:ug:chordless}). The moral graph of \ref{fig:dag:v-struct} is
	\ref{fig:dag:moral}, which in fact is another chordal cover of
	\ref{fig:ug:chordless}, and thus none of its orientations will be Markov
	equivalent to \ref{fig:ug:dageq}.
\end{exm}

\section{Gaussian parametrization}\label{sec:gparam}
When restricting to multivariate Gaussian distributions, we find connections
between conditional and vanishing parameters. This correspondence can be used
for providing a direct interpretation of Markov properties, both in the
undirected and directed case, allowing an enhanced manipulation of these Markov
models.

In the following, the elements of a real $q\times r$ matrix $\mat{M} \in
\mgens{q}{r}{\sreal}$ will be denoted as $m_{ij}$, where $i \in \set{1, \ldots,
q}$ and $j \in \set{1, \ldots, r}$. $\mat{M}_{IJ}$ will be the
$\card{I}\times\card{J}$ sub-matrix of $\mat{M}$, where
$I\subseteq\set{1,\ldots,q}$ and $J\subseteq\set{1,\ldots,r}$; and we will use
$\finv{\mat{M}}_{IJ}$ as $\finv{(\mat{M}_{IJ})}$. $\msympdefs$ and $\msymspdefs$
will represent the sets of positive and semi-positive definite symmetric
matrices, respectively. The $p$-variate Gaussian distributions is denoted as
$\dmnorm{p}{\vmean}{\mcov}$, where $\vmean \in \sreal^p$ is the mean vector and
$\mcov \in \msympdefs$ is the covariance matrix. $\midendim{p}$ will denote the
identity $p$-dimensional square matrix, whereas $\viden{p}$ will denote the
{$p$-}vector with all entries equal to $1$; many times, dimensionality
sub-scripts will be dropped if the dimension of the respective object is clear
from the context.

\subsection{Conditional independence and the multivariate Gaussian distribution}

Let $V = \set{1, \ldots, p}$. When a random vector $\vrand$ is distributed as
$\dmnorm{p}{\vmean}{\mcov}$, then for $i, j \in V$, $X_i \sindep X_j$ is
equivalent to $\sigma_{ij} = 0$. If we consider a partition $(I,J)$ of $V$, then
$\vrand_{I} \sgiven \vecb{x}_J$ is distributed as
$\dmnorm{\card{I}}{\vmean_I}{\mcov_{\syule{I}{J}}}$, where $\mcov_{\syule{I}{J}}
= \mcov_{II} - \mcov_{IJ}\finv{\mcov}_{JJ}\mcov_{JI}$ \citep{anderson2003}.
Thus, for $i, k \in I$, we have that $\ci{X_i}{X_k}{\vecb{x}_J}$ is equivalent
to $\scov_{\syule{ik}{J}} = 0$, the $(i, k)$ element in the conditional
covariance matrix $\mcov_{\syule{I}{J}}$. 

A correspondence can be established between the zeros in $\mcov_{\syule{I}{J}}$
and zero patterns in other representative matrices
\citep[][\S9.1]{wermuth1976,wermuth1980,uhler2018b}, as follows. Let the concentration matrix of
$\vrand$ be $\mconc = \finv{\mcov}$, with elements $\sconc_{uv}$ for $u, v \in
V$. The matrix $\mcov_{IJ}\finv{\mcov}_{JJ}$ is usually denoted as
$\mcoef_{\syule{I}{J}}$ and called the matrix of \textit{regression
coefficients} of $\vrand_{I}$ on $\vrand_{J}$. Letting $\mconc_{\syule{I}{J}}
\sdefeq \finv{\mcov}_{JJ}$, we have the following matrix identity
\citep{horn2012} 
\[
\mconc = 
	\finv{
		\begin{pmatrix}
			\mcov_{II}&\mcov_{IJ}\\
			\mcov_{JI}&\mcov_{JJ}
		\end{pmatrix}} =
	\begin{pmatrix}
		\finv{\mcov}_{\syule{I}{J}} & -\finv{\mcov}_{\syule{I}{J}}\mcoef_{\syule{I}{J}} \\
		- \mcoef_{\syule{I}{J}}^t\finv{\mcov_{\syule{I}{J}}} & \mconc_{\syule{I}{J}} + \mcoef_{\syule{I}{J}}^t\finv{\mcov}_{\syule{I}{J}}\mcoef_{\syule{I}{J}}
	\end{pmatrix}.
\]
This allows us to relate $\mcov_{\syule{I}{J}}$ with $\mconc$ and
$\mcoef_{\syule{I}{J}}$ as
\begin{align}
	&\mcov_{\syule{I}{J}} = \finv{\mconc}_{II},\label{eq:mconc:iden}\\
	&\mcoef_{\syule{I}{J}} = - \finv{\mconc}_{II}\mconc_{IJ},\label{eq:coeff:iden}
\end{align}
{which implies that}, dually, $\mconc_{II}$ {is identically equal to} the
concentration matrix of $\vrand_{I} \sgiven \vecb{x}_{J}$, while
$\mconc_{\syule{I}{J}}$ {is} the concentration matrix of $\vrand_{J}$. Finally,
from \eqref{eq:mconc:iden} we get, for $i, k \in V$,
\begin{equation}\label{eq:ug:mvn:indep}
	\ci{X_i}{X_j}{\vrand_{V\setminus\set{i, k}}} \iff \sconc_{ik} = 0,
\end{equation}
whereas {from \eqref{eq:coeff:iden} it follows that,} for $J \subseteq V$, $i, k \in V\setminus J$,
\begin{equation}\label{eq:dag:mvn:indep}
	\ci{X_i}{X_k}{\vrand_{J}} \iff \scoef_{\syule{ik}{J{\cup\set{k}}}} = 0,
\end{equation}
where $\scoef_{\syule{ik}{J{\cup\set{k}}}}$ is the $v$ entry in the vector
$\vcoef_{\syule{i}{J \cup \set{k}}}^t$, that is, the coefficient of $X_{k}$ on
the regression of $X_i$ on $\vecb{x}_{J \cup \set{k}}$. The original notation
for this, introduced in \citet{yule1907}, was $\scoef_{\syule{ik}{J}}$; that is,
$k$ is implicitly considered as included in the conditioning indexes. We have
however chosen the alternative, explicit notation
$\scoef_{\syule{ik}{J{\cup\set{k}}}}$, since it provides more notational
simplicity in later sections.

\subsection{Gaussian Markov models}

In the Gaussian case, undirected Markov models are in correspondence with the
concentration matrix, while for acyclic digraphs this correspondence is with the
regression coefficients. Both rely on the auxiliary Markov properties that we
presented in Section \ref{sec:mp}.

Let $\gdesc{\graph{G}}{V}{E}$ be an undirected graph and consider $\vrand$
distributed as $\pdist{\vrand} \equiv \dmnorm{p}{\vmean}{\mcov}$ with
$\pdist{\vrand} \in \gmm{M}{\graph{G}}$. Since $\pdist{\vrand}$ is globally
$\graph{G}$-Markov, it is also pairwise $\graph{G}$-Markov, and thus
\eqref{eq:ug:mvn:indep} directly gives that $\sconc_{uv} = 0$ for all $u, v \in
V$ such that $uv \in E$. This means that, if we define the set
$\mugmm{\graph{G}} \sdefeq \set{\mat{M}\in\msympdefs \st m_{uv} = 0 \text{ for
all } uv \notin E}$, we have $\mconc \in \mugmm{\graph{G}}$ if and only if
$\pdist{\vrand}$ is pairwise $\graph{G}$-Markov. Furthermore, since the
multivariate Gaussian distribution has positive density, $\firel{I}{\vrand}$ is
a graphoid and thus the three Markov properties are equivalent. This allows us
to redefine the \textit{Gaussian undirected Markov model} as
\begin{equation}\label{eq:ggmm:ugdefn}
	\ggmm{\graph{G}} = \set{\dmnorm{p}{\vmean}{\mcov} \st \finv{\mcov} \in \mugmm{\graph{G}}{, \vmean \in \sreal^p}}.
\end{equation}

In the directed case, the redefinition is not so direct. Let
$\gdesc{\graph{D}}{V}{A}$ be an acyclic digraph, and assume, for notational
simplicity, that the nodes are already ancestrally ordered as $1 \preceq \cdots
\preceq p$. If $\vrand$ is distributed as $\pdist{\vrand} \equiv
\dmnorm{p}{\vmean}{\mcov}$ with $\pdist{\vrand} \in \gmm{M}{\graph{D}}$, it
satisfies the ordered Markov property. Thus, whenever $v \in
\pr(u)\setminus\pa(u)$, we have $\ci{X_u}{X_v}{\vrand_{\pa(u)}}$, which is
equivalent to $\scoef_{\syule{uv}{\pa(u){\cup\set{v}}}} = 0$ as in
\eqref{eq:dag:mvn:indep}. Since we have assumed an ancestral order,
$\scoef_{\syule{uv}{\pa(u){\cup\set{v}}}} = \scoef_{\syule{uv}{\pr(u)}}$ for all
$u \in V$, $v \in \pr(u) \setminus \pa(u)$, which leads to $\pdist{\vrand}$
being ordered $\graph{D}$-Markov if and only if $\scoef_{\syule{uv}{\pr(u)}} =
0$ for all $u \in V$, $v \in \pr(u) \setminus \pa(u)$. Such triangular
requirement on the regression coefficients can be expressed with the matrix
$\mcoef$ defined, for $v < u $ as $b_{uv} = 0$ if $v \notin \pa(u)$, and $b_{uv}
= \scoef_{\syule{uv}{\pa(u)}} \equiv \scoef_{\syule{uv}{\pr(u)}}$ otherwise. 

If we let $\sccov_u \sdefeq \scov_{\syule{uu}{\pr(u)}}$, the previous
characterization leads to a matrix form of the linear regressions involved as
\begin{equation}\label{eq:mlr}
\vrand = \vmean + \mcoef(\vrand - \vmean) + \vnoise,
\end{equation} 
where $\snoise_u \sdist
\dnorm{0}{\sccov_u}$. We can rearrange it as $\vrand = \finv{\mdec}\vconst +
\finv{\mdec}\vnoise$,
where $\vconst \sdefeq \mdec\vmean$ and $\mdec \sdefeq
\midendim{p} - \mcoef${, since $\mdec$ is invertible}. {Let} $\mccov$ be the
diagonal matrix of conditional variances $\vccov$. {Sometimes $\vconst$,
$\mcoef$ and $\mccov$ are called the $\graph{D}$-parameters of $(\vmean, \mcov)$
\citep{andersson1998}. In fact,} $\mdec$ and $\mccov$ allow a decomposition of
$\mcov$ (and $\finv{\mcov}$) as $\mcov = \finv{\mdec}\mccov\mdec^{-t}$.
Furthermore, this decomposition uniquely determines $\mcov$ via $\mdec$/$\mcoef$
and $\mccov$ \citep{horn2012}. Thus, in analogy with \eqref{eq:ggmm:ugdefn}, if
we define the set 
\[
	\mdgmm{\graph{D}} \sdefeq \set{\mat{M}\in\mgens{p}{p}{\sreal}
\st m_{uv} = 0 \text{ for all } (u, v) \notin A}
\] 
and the set $\mdiags{p}$ of $p
\times p$ diagonal matrices, we can redefine the \textit{Gaussian directed
Markov model} as
\begin{equation}\label{eq:ggmm:dagdefn}
	\ggmm{\graph{D}} = \set{\dmnorm{p}{\vmean}{\mcov} \st \finv{\mcov} = (\midendim{p} - \mcoef)^t\finv{\mccov}(\midendim{p} - \mcoef),\, \mcoef \in \mdgmm{\graph{D}}{,\, \mccov \in \mdiags{p}}}.
\end{equation}

\section{Maximum likelihood estimation}\label{sec:mle}
Maximum likelihood estimation is greatly simplified in exponential family theory
\citep{barndorff-nielsen1978}. The multivariate Gaussian distribution is a
regular exponential family, and thus both undirected and directed Gaussian
Markov models can be expressed as special subfamilies of it.

\subsection{The Gaussian family and maximum likelihood}
In the multivariate Gaussian family the canonical parameter is $\fecan =
(\mconc\vmean, {-}\mconc{/2})$, over the space {$\ecans = \{(\fecan_1, \fecan_2)
\st \fecan_1 \in \sreal^p, -\fecan_2 \in \msympdefs\}$} and the sufficient
statistics are $\fesuf{{\vrand}} = ({\vrand}, {\vrand}{\vrand^t})$. Let
$\{\vsreal{n}\st 1\leq n\leq N\}$ be $N$ independent observations, where
$\vsrand{n}\sdist\dmnorm{p}{\vmean}{\mcov}$ for each $n\in\set{1,\ldots,N}$,
arranged in $\msample \in \mgens{p}{N}{\sreal}$, the respective random matrix
being $\msrand$. The random sample is also a regular exponential family with
canonical parameter $\fecan = (\mconc\vmean, -\mconc/2)$ over the space
$\ecans$. The sufficient statistics in this case are $\fesuf{{\msrand}} =
(N{\vsmrand}, {\msrand}{\msrand}^t)$ with $N{\vsmrand} =
\sum_{n=1}^N\vsrand{n}$.

In a regular exponential family $\fexp$, a maximum of the likelihood function,
$\flike{\fecan}$, given a random sample $\msrand = \msample$, is reached in
$\ecans$ if and only if $\fesuf{\msample}$ {belongs to the interior of}
$\scsup{\sesuf}$, the {closed} convex hull of the support of the {distribution
of} $\sesuf${, denoted as $\sint(\scsup{\sesuf})$}. In such case, it is unique
and given by the $\fecan \in \ecans$ satisfying $\expec{\fesuf{\msrand}} =
\fesuf{\msample}$.

For the multivariate Gaussian random sample, we have that $\expec{N{\vsmrand}} =
N\vmean$ and $\expec{{\msrand}{\msrand}^t} = N\mcov + N\vmean\vmean^t$, thus the
convex support of $\fesuf{{\msrand}} = (N{\vsmrand}, {\msrand}{\msrand}^t)$ is
$\scsup{\sesuf} = \{({\vecb{v}}, \mat{M}) \in \sreal^p\times\msyms{p} \st
\mat{M} - {\vecb{v}\vecb{v}}^t/N \in \msymspdefs\}$. This gives that the maximum
likelihood estimator for $(\vmean, \mcov)$ exists if and only if
$\msample\msample^t - N\vsmean\vsmean^t \in \msympdefs$, which happens with
probability one whenever $N > p$ and never otherwise. The solution in such case
is $(\vsmean, \msprod/N)$, where 
\[
	\msprod = \sum_{n = 1}^N\pgroup{\vsrand{n} - {\vsmrand}}\pgroup{\vsrand{n} - {\vsmrand}}^t {= {\msrand}{\msrand}^t - N{\vsmrand}{\vsmrand}^t}.
\]

A particular situation, usually assumed, is when $\vmean = {\vzero}$. The
canonical parameter now is $\fecan = -\mconc/2$ in the space $\{\fecan \st
-\fecan \in \msympdefs\}$, and the sufficient statistic {is} $\fesuf{{\msrand}}
= {\msrand}{\msrand}^t$. The maximum likelihood estimator exists if and only if
$\msample\msample^t \in \msympdefs$, and in such case it is
${\msrand}{\msrand}^t/N = \msprod/N$.

\subsection{Gaussian Markov models as exponential families}
When $\graph{G}$ is an undirected graph, the set $\mugmm{\graph{G}}$ is a convex
(linear) cone inside the positive definite cone $\msympdefs$
{\citep[e.g.][\S9.2]{uhler2018b}}, which means that $\sreal^p\times\mugmm{\graph{G}}$ is an
affine subspace of $\sreal^p\times\msympdefs$, and thus $\ggmm{\graph{G}}$ is
also a regular exponential family \citep{barndorff-nielsen1978}. 
Assume that $\vmean = {\vzero}$ and let
$\mumat{\msprod}{\graph{G}}$ be the projection of $\msprod$ on
$E \cup \set{uu \st u \in V}$, that is, such that
$q^{\graph{G}}_{uv} = 0$ for all $uv \notin E$ with $u \neq
v$. Since $\flike{\mconc} \propto
\det(\mconc)^{\frac{1}{2}}\exp\pgroup{-\tr(\mconc\msprod)}$
and $\mconc \in \mugmm{\graph{G}}$, we have
$\tr(\mconc\msprod) = \tr(\mconc\mumat{\msprod}{\graph{G}})$
and the sufficient statistic for $\ggmm{\graph{G}}$ is
$\fesuf{\msample} = \mumat{\msprod}{\graph{G}}$
\citep{lauritzen1996}. Its convex support is $\scsup{\sesuf} =
\{\mumat{\mat{P}}{\graph{G}} \st \mat{P} \in \msymspdefs\}$,
equivalently called the set of projections extendible to full
positive definite matrices. Thus, the maximum likelihood
estimator for $\mcov$ exists if and only if
$\mumat{\msprod}{\graph{G}} \in \sint(\scsup{\sesuf})$, which
is equivalent to say that $\mumat{\msprod}{\graph{G}}$ is
extendible to a full positive definite matrix. Whenever it
exists, it is the only extendible matrix $\egen{\mcov}$ that
also satisfies the model restriction $\finv{\egen{\mcov}} \in
\mugmm{\graph{G}}$. A sufficient condition thus is that
$\msprod \in \msympdefs$, which happens almost surely for $N
\geq p$. {Recovering $\egen{\mcov}$ is a convex optimization
problem, \citet{uhler2018b} \S9.6 overviews some of the algorithms available for its
computation. Note however that if $\graph{G}$ is chordal, then there is a closed
form expression for $\egen{\mconc}$ \citep{lauritzen1996}}

{The existence of $\egen{\mcov}$ has been
completely characterized 
when $\graph{G}$ is chordal by \citet{grone1984} and \citet{frydenberg1989},
independently. Since finding
$\egen{\mcov}$ is equivalent to a positive definite matrix completion problem
\citep[][\S9.3]{uhler2018b}, the problem lies at the interface between
statistic{s} and linear algebra. Therefore, this problem has been solved
from an
algebraic \citep[][\S9.4]{sturmfels2010,uhler2012,uhler2018b} viewpoint for a
general, non-chordal $\graph{G}$.
However, the conditions on the 
sample size $N$ are still unknown except for certain non-chordal graph types, see
\citet{uhler2018b} \S9.5 for an up-to-date overview of the advances
made so far.}

{Now we turn on the case where the random sample $\msrand$ is assumed to a
follow multivariate Gaussian distribution constrained by the separation
properties in an acyclic digraph. The restriction in \eqref{eq:ggmm:dagdefn},
however, is not linear in the canonical parameter; in fact, 
\citet{spirtes1997} show that they are curved exponential families. To
obtain the maximum likelihood estimates, theory from multivariate linear
regression can be applied \citep{andersson1998}. Recall that if $\vrand \sim
\dmnorm{p}{\vmean}{\mcov}$ and $\dmnorm{p}{\vmean}{\mcov} \in \ggmm{\graph{D}}$,
then $\vrand$ can be expressed as \eqref{eq:mlr}. Thus, we can estimate the
$\graph{D}$-parameters for $(\vmean, \mcov)$ as the usual least squares
estimators,}
\begin{align*}
	&\egen{\vcoef}_{\syule{u}{\pa(u)}}^t = \msprod_{u\pa(u)} \finv{\msprod_{\pa(u)\pa(u)}},\\
	&{\egen{\sconst}_u = \ssmean_u - \egen{\vcoef}_{\syule{u}{\pa(u)}}^t\vsmean_{\pa(u)},}\\
	&N\egen{v}_{uu} = q_{uu} -
	\egen{\vcoef}_{\syule{u}{\pa(u)}}^t\msprod_{u\pa(u)}^t,
\end{align*}
respectively for each $u \in V${, where $q_{uu}$ is the $u$-th
diagonal entry in $\msprod$}. {We can then obtain directly the maximum
likelihood estimator for $(\vmean, \mcov)$ from their respective
$\graph{D}$-parameter estimators \citep[see][for an algorithm]{andersson1998}.
As opposed to the undirected case, $(\egen{\vmean}, \egen{\mcov})$} exist with
probability one if and only if $N \geq p + \max\set{\card{\pa(u)} \st u \in V}$
\citep{anderson2003}. 
{Recently, \citet{bendavid2012} analyze in detail the relationship
between the $\graph{D}$-parameters and $\mcov$ as a positive definite matrix
completion problem, in analogy with the undirected case.}

\section{{Model selection via} hypothesis testing}\label{sec:hypo}
Maximum likelihood estimators, presented in the previous section, can be used to
address the problem of model estimation, and require either prior knowledge or a
statistical procedure that allows model selection{; that is, selecting the graph
that will define the Markov model}. In this section we will review the main
hypothesis testing methods for such task. 

{Throughout the section, for $u, v \in V = \{1, \ldots, p\}$ and $U
\subseteq V \setminus \{u, v\}$, we will denote as $\rho_{uv \cdot U}$ the
partial correlation coefficient between $u$ and $v$ given the variables in $U$,
and as $r_{uv\cdot U}$ its maximum likelihood estimator, the sample partial
correlation \citep[see e.g.][\S4.3 for an introduction to partial correlation
theory]{anderson2003}.}

\subsection{Stepwise selection}

In the undirected case, we are interested in testing the hypothesis
$H_0:\mconc\in\mugmm{\graph{G}_0}$ against $H{_1}:\mconc\in\mugmm{\graph{G}}$,
where $\gdesc{\graph{G}_0}{V}{\sedges{E}{\graph{G}_0}} \subseteq
\gdesc{\graph{G}}{V}{\sedges{E}{\graph{G}}}$. {The result of such test
determines whether the edges in $\sedges{E}{\graph{G}} \setminus
\sedges{E}{\graph{G}_0}$ should be excluded from the selected model; that is why
these tests are usually known as \textit{edge exclusion tests}. Note also that
this is \textit{backward} model selection, since our null hypothesis consists on
a subgraph.} Let $\egen{\mcov}_0$ and $\egen{\mcov}$ be the maximum likelihood
estimators for {a covariance matrix in the Markov model determined by}
$\graph{G}_0$ and $\graph{G}${, respectively}. The likelihood ratio statistic is
\[	
\etest{{L}} =
\frac{\det(\egen{\mcov})^{N/2}}{\det(\egen{\mcov}_0)^{N/2}}
= \left(\frac{\det(\egen{\mconc}_0)}{\det(\egen{\mconc})}\right)^{N/2}.
\]

Under $H_0$, $-2\log(\etest{{L}}) = N(\log\det(\egen{\mconc}) -
\log\det\egen{\mconc}_0)$ is asymptotically distributed as a $\chi^2$
distribution with $\card{\sedges{E}{\graph{G}}}-\card{\sedges{E}{\graph{G}_0}}$
degrees of freedom; however, this is a poor approximation in many cases
\citep{porteous1989}. More accurate distributional results have been derived by
\citet{eriksen1996}, as follows. Let $\graph{G}_0\subset \ldots\subset
\graph{G}_k (= \graph{G})$ be a sequence of graphs where, for $1\leq i\leq k$,
$\sedges{E}{\graph{G}_{i-1}} = \sedges{E}{\graph{G}_i}\setminus\set{e_i}$ for
some $e_i=u_iv_i\in \sedges{E}{\graph{G}_i}$ (sequence of edge deletions).
{Then,
under $H_0$, $\etest{{L}}^{2/N}$ is distributed as the
product $\prod_{i=1}^kB_i$ of univariate Beta variables, where, for $1\leq i\leq
k$,
\[
	B_i \sdist \dbeta{\frac{1}{2}(N -
	\card{\nei_{\graph{G}_i}(u_i)\cap\nei_{\graph{G}_i}(v_i)} - 2)}{\frac{1}{2}}.
\]
The above result is exact whenever $\graph{G}$ and $\graph{G}_0$ are chordal or
share the same non-chordal maximal subgraphs \citep{eriksen1996}.
{Specifically, denote as $C_i^* = \nei(u_i)\cap\nei(v_i)$ the unique
clique in $\graph{G}_i$ of which edge $e_i$ is a member. Then, under $H_0$
\citep[see e.g.][Proposition 5.14]{lauritzen1996}
\begin{equation*}
\etest{{L}}^{2/N} { = \prod_{i = 1}^k(1 -
r_{u_iv_i\cdot C_i^*\setminus\{u_iv_i\}}^2)},
\end{equation*}
giving that $T_L^{2/N}$ is distributed as $\prod_{i = 1}^k\mathcal{B}((N -
|C_i^*|)/2, 1/2)$.
Note that in this decomposable case one avoids to actually
compute $\egen{\mconc}$ and $\egen{\mconc}_0$. The statistic $T_L$ has been used for model
selection in undirected Gaussian Markov models by \citet{wermuth1976b}.}

In the case of a directed Gaussian Markov model over an acyclic digraph
$\gdesc{\graph{D}}{V}{A}$, most of the results are adaptations from analogues in
multivariate linear Gaussian models. The likelihood ratio, whose moments are
also characterized in \citet{andersson1998}, is
\begin{equation*}	
	\etest{{L}} =
	\frac{\det(\egen{\mcov})^{\frac{N}{2}}}{\det(\tilde{\mcov})^{\frac{N}{2}}} =
	\frac{\prod_{v \in V}\abs{\egen{\scov}_{vv} -
	\egen{\vcov}_{\syule{v}{\pa(v)}}^t\finv{\egen{\mcov}}_{\pa(v)}\egen{\vcov}_{\syule{v}{\pa(v)}}}}{\prod_{v
	\in V}\abs{\tilde{\scov}_{vv} -
	\tilde{\vcov}_{\syule{v}{\pa(v)}}^t\finv{\tilde{\mcov}}_{\pa(v)}\tilde{\vcov}_{\syule{v}{\pa(v)}}}},
\end{equation*}
where $\tilde{\mcov}$ and $\egen{\mcov}$ are the respective maximum likelihood
estimators for $\tilde{\graph{D}}$ and $\graph{D}$, $\tilde{\graph{D}} \subseteq
\graph{D}$. 

{
A backward stepwise method has become popular for
selecting $\graph{D}$, commonly called the \textit{PC algorithm}
\citep{spirtes2000}. This method proceeds by first finding an estimator of the
skeleton,
$\egen{\gskel{\graph{D}}}$, from a complete undirected graph, and then
orienting it. At iteration $i$ of the first step (finding $\egen{\gskel{\graph{D}}}$), $H_0 : \ci{X_u}{X_v}{X_C}$ is tested, with
$C = \egen{\nei}(u) \setminus \{v\}$ and $\card{C} = i$. The edge
$uv$ is removed from $\egen{\gskel{\graph{D}}}$ if $H_0$ is not rejected. Note
that
$\egen{\gskel{\graph{D}}}$ depends on the order in which $H_0$ is tested at each
iteration, problem circumvented in the modification by \citet{colombo2014}.
Assuming that $\firel{I}{\vrand} = \firel{I}{\graph{D}}$ (see Section
\ref{sec:markov}), commonly called the \textit{faithfulness} assumption,
\citet{robins2003} showed that the PC algorithm is pointwise consistent but may
not be uniformly consistent, regardless of the method used for testing $H_0$.
\citet{zhang2003} approached this problem by introducing a stronger condition,
called \textit{strong faithfulness}, which, by requiring nonzero partial
correlations to have a common lower bound, gives uniform consistency, even in a
high-dimensional setting \citep{kalisch2007}. However, despite the set of
`unfaithful' distributions has Lebesgue measure zero \citep{meek1995}, those
`strongly unfaithful' constitute a non-zero Lebesgue measure set, which can in
some cases be very large \citep{uhler2013,lin2014}.}

\subsection{Multiple testing}
When performing model selection with these tests, multiple testing error rates
{need to be} controlled. For overcoming this, \citet{drton2004} {propose an
alternative to the previous stepwise methods.}
{
First, note that both acyclic directed and undirected Gaussian Markov models
over $V = \{1, \ldots, p\}$ are
characterized by certain partial correlation coefficients, since for $u, v \in
V$ and $U \subseteq V \setminus \{u, v\}$, we have $\ci{X_u}{X_v}{\vecb{X}_U}
\iff \rho_{uv \cdot U} = 0$. Assuming conditional independence, that is, $\scorr_{\syule{uv}{U}} = 0$, then
$\sqrt{N - \card{U} - 2} \, r_{\syule{uv}{U}} / \sqrt{1 - r_{\syule{uv}{U}}^2}$
has a $t$ distribution with $N - \card{U} - 2$ degrees of freedom. However, a
faster Gaussian approximation can be obtained using Fisher's $Z$-transform,
\[
	Z(x) = \frac{1}{2}\log\pgroup{\frac{1 + x}{1 - x}} = \finv{\tanh}(x).
\]
In such case the distribution of $\sqrt{N - |U| - 3}\,Z(r_{uv\cdot U})$ tends to a standard Gaussian.}

{Based on the above discussion, \citet{drton2004} propose a method
where} a set of simultaneous $p$-values
and confidence intervals is obtained {such that} the edge set is estimated, for
a significance level $\alpha$ and using \citet{sidak1967} inequality, as
\begin{equation}\label{eq:ht:multiple:ug}
	\egen{E}^\alpha \sdefeq \set{uv :
	\sqrt{N-p - 1}\abs{{Z(r_{uv \cdot V \setminus \{u,
v\}})}} >
	\finv{\fdstand}\pgroup{\frac{1}{2}(1-\alpha)^{\frac{2}{p(p-1)}} +
	\frac{1}{2}}},
\end{equation}
where {$\fdstand$ is the cumulative distribution function of a standard
Gaussian}. Denoting as $\gdesc{\egen{\graph{G}}^\alpha}{V}{\egen{E}^\alpha}$, it
holds $\liminf_{N\tendsto\infty}P(\egen{\graph{G}}^\alpha = \graph{G})\geq 1
-\alpha$ if the distribution under consideration $\dmnorm{p}{\vmean}{\mcov} \in
\ggmm{\graph{G}}$ is faithful to $\graph{G}$, that is, if $\sconc_{uv} = 0 \iff
uv \notin E$. If faithfulness is not satisfied, then the result hold{s} with
respect to the {smallest} graph {$\graph{H}$ such that} $\graph{G} \subseteq
\graph{H}$ and {$\dmnorm{p}{\vmean}{\mcov}$ is faithful to $\graph{H}$}.

The multiple testing procedure in \eqref{eq:ht:multiple:ug} has also
been extended in \citet{drton2008}, obtaining an estimate of the arc set as
\begin{equation}\label{eq:ht:multiple:dag}
	\egen{A}^\alpha \sdefeq \set{(v, u) : v < u \text{ {and} } \sqrt{N - u -
	1}\abs{{Z(r_{uv \cdot \pr(u) \setminus \{v\}})}} >
	\finv{\fdstand}\pgroup{\frac{1}{2}(1-\alpha)^{\frac{2}{p(p-1)}} +
	\frac{1}{2}}},
\end{equation}
where an ancestral ordering $\prec$ is being assumed in $V$ such that the
resulting permutation is the identity; that is, such that $v \prec u \iff v <
u$. Consistency is established as in the undirected case; note the symmetry with
\eqref{eq:ht:multiple:ug}. See \citet{drton2007} for a general discussion on
some variations of \eqref{eq:ht:multiple:ug} and \eqref{eq:ht:multiple:dag} and
their impact on overall error control. {Recently, \citet{liu2013}
has extended the methodology of \citet{drton2004} to the high dimensional
scenario, with $p > N$.}

{ 
A related testing procedure has emerged motivated by the field of gene network
learning from microarray data. Instead of testing full partial correlations
$\rho_{uv \cdot V \setminus \{u, v\}}$ in an undirected model, only limited $q$-order partial
correlations $\rho_{uv \cdot U} = 0$, where $U \subseteq V \setminus \{u, v\}$
and $|U| = q$, are tested \citep{castelo2006}. An edge is added to the resulting
graph, called \textit{$q$-partial graph}, only when all of the $q$-partial
correlations are rejected to be zero. This procedure is specially suited for
situations where the number of variables is substantially larger than the number
of instances, as happens in the case of microarray data, where low order
conditional independence relationships (up to $q \leq 3$) have been popular
\citep{wille2006,fuente2004,magwene2004}. \citet{castelo2006} generalize and
formalize these approaches, and provide a robust model selection procedure for
$q$-partial graphs. This is intended to serve as an intermediate step for model
selection of a classical undirected Markov model $\ggmm{\graph{G}}$, and yields
to a great simplification when $\graph{G}$ is sparse \citep{castelo2006}.
}

\section{Regularization}\label{sec:reg}
Regularization approaches, which perform model selection and
estimation in a simultaneous way, have become popular in the
context of Markov models. They are usually applied when $N <
p$, and thus the existence of the maximum likelihood estimator
is not guaranteed. The main consistency results available for
both the directed and undirected case{s} share sparseness and
high-dimensionality assumptions, as we will see {below}. There
are two different approaches, those that penalize the
likelihood and those that instead focus on the regression
coefficients.

{Througout this section, we will employ the asymptotic
notation, specifically symbols $\bigo{\cdot}$ and
$\bigtheta{\cdot}$, asymptotic inferiority and equivalence,
respectively.} {For $\mat{M} \in \mgens{q}{r}{\sreal}$},
$\vect(\mat{M})$ will {denote} the vectorized function of
$\mat{M}${, $(m_{11}, \ldots, m_{q1}, \ldots, m_{1r}, \ldots,
m_{qr})^t$}. This way, the operator norm of $\mat{M}$ will be
denoted as $\norm{\mat{M}}$; whereas $\norm{\mat{M}}_{{q +
r}}$ will be used to denote $\norm{\vect(\mat{M})}_{{q + r}}$,
being $\norm{\cdot}_p$ the $p$-norm function. If {$\vecb{v}$}
is a $p$-vector, $\diag({\vecb{v}})$ will denote the matrix
$\mat{M}$ {in $\mdiags{p}$} with {main} diagonal ${\vecb{v}}$;
analogously, $\diag(\mat{M}) \in \mdiags{p}$ will {have} the
same diagonal as $\mat{M}${, and} $\mat{M}^-$ will be used for
$\mat{M} - \diag(\mat{M})$. 

\subsection{Node-wise regression}
{Let $\gdesc{\graph{G}}{V}{E}$ be an undirected graph, with $V
= \set{1, \ldots, p}$. 
{
Let $X$ be a random vector whose distribution
$\dmnorm{p}{\vmean}{\mcov}$ belongs to the undirected Gaussian
Markov model $\ggmm{\graph{G}}$. Assume, for notational
simplicity, that $\vmean = \vzero$, and, following the
notation of Section \ref{sec:mle},
let $\msrand = \msample$ be a $p \times N$ random sample from
$\dmnorm{p}{\vzero}{\mcov}$. Since for each $u, v \in V$,
$\scoef_{\syule{uv}{V\setminus\set{u}}} = -
\sconc_{uv}/\sconc_{uu}$ (Equation \eqref{eq:coeff:iden}),
then
\[
	\ci{X_u}{X_v}{\vrand_{V \setminus \set{u, v}}} \iff
	\sconc_{uv} = 0 \iff \scoef_{\syule{uv}{V\setminus
	\set{u}}} = \scoef_{\syule{vu}{V\setminus \set{v}}} = 0.
\]
This means that an analogue of the matrix $\mcoef$ in directed
Gaussian Markov models (Equation \eqref{eq:mlr}) can be used
for determining the missing edges in the undirected case.}
In \citet{meinshausen2006},
{this} is done in the regression function, as}
\begin{equation}\label{eq:ug:reg:reg}
	\egen{\vecb{b}}_u^\slagr \sdefeq \argmin{\vecb{b}_u \in
	\sreal^p, \,b_{uu} =
	0}{\frac{1}{N}\norm{\msample_{u}^t -
	\msample^t\vecb{b}_u}_2^2 + \slagr f(\vecb{b}_u)},
\end{equation}
where $\slagr\geq 0$, {$\msample_{u}$ is the
$u$-the row vector of $\msample$} {and $f(\cdot)$ is the
penalty function}. {For each $v \in V \setminus
\{u\}$,
$\egen{b}_{uv}^\slagr$ gives an estimate of
$\scoef_{\syule{uv}{V\setminus \set{v}}}$.}
{Let} $\egen{\nei}(v) \sdefeq \{u\in V:
\egen{b}^\slagr_{vu}\neq 0\}$. {
while} $u \in \nei(v) \iff v \in \nei(u)$ for all $u, v \in
V$, this
may not be true {for $\egen{\nei}(u)$ and
$\egen{\nei}(v)$. Hence, two different estimators for
the edge set $E$ may be defined}
\begin{align*}
	&\egen{E}_{\wedge} \sdefeq \set{uv: u\in\egen{\nei}(v)\text{ and }v\in\egen{\nei}(u)},\\
	&\egen{E}_{\vee} \sdefeq \set{uv: u\in\egen{\nei}(v)\text{ or }v\in\egen{\nei}(u)}.
\end{align*}
Let $f(\cdot) =
\norm{\cdot}_1$, commonly known as the \textit{lasso} penalty
\citep{tibshirani1996} {or $l_1$ regularization}. Then both
estimators {$\egen{E}_\wedge$ and $\egen{E}_\vee$} are consistent
for certain choice of $\lambda$. {This result was independently
discovered by \citet{meinshausen2006}, \citet{zhao2006}, \citet{zou2006} and
\citet{yuan2007b}. It relies on the following almost necessary and sufficient
condition}
\begin{equation}\label{eq:reg:irrep}
\abs{\sum_{z\in\nei(v)}\sign(\scoef_{\syule{vz}{\nei(v)}})\scoef_{\syule{uz}{\nei(v)}}}
< 1,
\end{equation}

{This node-wise regression approach may also be used to perform
model selection for acyclic directed Gaussian Markov models if there is a known
order among the variables, see for example \citet{shojaie2010} or \citet{yu2017}
and references therein. From Equation \eqref{eq:mlr}, the regression function
to penalize in this case would be, for each $u \in V$, 
\begin{equation}\label{eq:dag:reg:reg}
	X_u = \mu_u + \sum_{v \in \pr(u)}\beta_{uv \cdot \pr(u)}(X_v - \mu_v) + E_u
\end{equation}
The condition of
Equation \eqref{eq:reg:irrep}, commonly called the `irrepresentable condition'
\citep{zhao2006} or `neighbourhood stability' \citep{meinshausen2006}, is
inherent to model selection in linear regression with $l_1$ regularization, and
thus it also holds when penalizing \eqref{eq:dag:reg:reg} with the $l_1$ penalty.
However, some variants have been proposed because it is rather restrictive.
These alternatives usually rely on thresholding the regression coefficients or
adding weights in the $l_1$ penalty, that under milder assumptions still achieve
model selection consistency
\citep{meinshausen2009} or other attractive, `oracle' properties
\citep{geer2009}; see \citet{buehlmann2011}, \S7 for a review. \citet{geer2009}
show that although model selection consistency for neighbourhood selection may
be restrictive, sufficient conditions for such oracle properties hold fairly
generally.
}

\subsection{Penalized likelihood}
In \cite{geer2013}, $l_0$ regularization is alternatively used in the context of
directed Gaussian Markov models{, without assuming a known order}.
As in \citet{meinshausen2006}, the regression coefficients are {penalized} in their
approach, more generally, the $\graph{D}$-parameters {in the
likelihood function (assuming $\vecb{\mu} = \vecb{0})$}. As such, the assumptions
required for the consistency of both methods share some symmetry, as we have
outlined in Table \ref{tab:sum:reg}. The estimators in this case are obtained as
\begin{equation*}
(\egen{\mccov}^\lambda, \egen{\mcoef}^\lambda) = \argmin{\mconc = (\midendim{p} -
\mcoef)^t\finv{\mccov}(\midendim{p} - \mcoef),\, \mcoef \in \mdgmm{\graph{D}},\,
\mccov \in \mdiags{p}}{\tr(\mconc\mscov)-
N\log\det(\mconc)  + \lambda{f(\mconc)}}, 
\end{equation*}
where $\lambda\geq 0${, $\mdgmm{\graph{D}}$, $\mdiags{p}$ are as
in Equation \eqref{eq:ggmm:dagdefn} and $\mat{S} = \mat{x}\mat{x}^t/N$. When $f(\mconc) = \card{\set{(u, v) \st
b_{uv} \neq 0}}$ ($l_0$ regularization), $\egen{\mccov}^\lambda$ and
$\egen{\mcoef}^\lambda$} are equal among Markov equivalent models {and the resulting estimator of the concentration matrix $\egen{\mconc}^\lambda$
is consistent for certain choice of $\lambda$} \citep{geer2013}.
{The strong faithfulness condition for the PC algorithm, bounding
nonzero partial correlations, resembles
the assumptions for regularization methods (Table \ref{tab:sum:reg}). In fact,
$l_0$ regularization has been suggested as an alternative for the PC, in order
to avoid the restrictive strong faithfulness assumption \citep{uhler2013};
however, it is unclear how the assumptions of both methods are related. For
recent extensions of the work by \citet{geer2013}, see \citet{aragam2015} and \citet{aragam2017}.}

\begin{table}
	\centering
	\footnotesize
	\begin{tabular}{l l}
		\toprule
		$\ggmm{\graph{G}}$ \citep{meinshausen2006} 	& $\ggmm{\graph{D}}$ \citep{geer2013} \\
		\midrule
		$l_1$ regularization & $l_0$ regularization \\
		Lower bound on $\abs{\scorr_{\syule{uv}{V\setminus\set{u, v}}}}$ & Lower
		bound on $\abs{\beta_{\syule{uv}{\pr(u)}}}$ \\
		Upper bound on $\card{\nei(v)}$ & Upper bound on $\card{\pa(v)}$ \\
		Bounded neighbourhood perturbations & Bounded permutation perturbations \\
		\bottomrule
	\end{tabular}
	\caption{\label{tab:sum:reg} Comparison of assumptions for consistency
	results on regression based penalized estimation in acyclic directed and undirected Gaussian Markov models.}
\end{table}

In {undirected Gaussian Markov models} conditional independences can
be {read} from $\mconc$. {Therefore, the penalized
likelihood approach can be formulated more
directly,} for $\slagr \geq 0$, as
\begin{equation}\label{eq:pen:like}
	\egen{\mconc}^\slagr = \argmin{\mconc \in \mugmm{\graph{G}}}{\tr(\mconc\mscov)-N\log\det(\mconc) + \slagr f(\mconc)}.
\end{equation}
\citet{yuan2007} {were} the first to pursue this approach, and they
chose $f(\mconc) = \norm{\mconc^-}_1${, that is, the off-diagonal
elements in $\mconc$, which determine the edges of the resulting undirected
graph, are penalized}. Later, in \citet{banerjee2008} the diagonal elements are
included in the regularization function, that is, $f(\mconc) =
\norm{\mconc}_1${; however, since $1/\sconc_{uu} = \scov_{\syule{uu}{V\setminus
\set{u}}}$, this choice for the penalty favours larger values for the error
variances in the regression of $X_u$ on the rest of variables
\citep{buehlmann2011}. Nonetheless, this estimator is the one chosen in the
extensively used algorithm \emph{Graphical Lasso} of \citet{friedman2008},
although model selection consistency has only been proved for $f(\mconc) =
\norm{\mconc^-}_1$ \citep{lam2009,ravikumar2011}. It is not known whether the
sufficient conditions required for this consistency are strictly stronger than
the irrepresentable condition, as some examples \citep{meinshausen2008} seem to
indicate.} 

{For the penalization of \citet{yuan2007} ($f(\mconc) =
\norm{\mconc^-}_1$)}, the convergence rate is
\citep{rothman2008}
\begin{equation*}
	\norm{\egen{\mconc}^\slagr - \mconc}_2 \in \bigo{\sqrt{\frac{(\card{E} + p)\log (p)}{N}}} \text{ as }N\tendsto\infty.
\end{equation*}
A relaxation of this rate can be obtained based on the correlation
matrix, as follows. Since $\mcov = \msd\mcorr\msd$ with $\mcorr$ the correlation
matrix and $\msd$ the diagonal matrix of standard deviations, if we let the
corresponding sample estimates be $\egen{\msd}^2 = {\diag}(\egen{\mcov})$ and
$\egen{\mcorr} = \finv{\egen{\msd}}\egen{\mcov}\finv{\egen{\msd}}$, {we can then
estimate} $\mat{K} = \finv{\mcorr}$ {as}
\[\egen{\mat{K}}^\slagr = \argmin{\mat{K} \in \mugmm{\graph{G}}}{\tr(\mat{K}\egen{\mcorr})-N\log\det(\mat{K})  + \slagr f(\mat{K})},\]
for $\slagr \geq 0$.
The concentration matrix can then be alternatively estimated as
$\vecb{\tilde{\mconc}}^\slagr =
\finv{\egen{\msd}}\egen{\mat{K}}^\slagr\finv{\egen{\msd}}$, yielding a
convergence rate of \citep{rothman2008}
\begin{equation*}
	\norm{\vecb{\tilde{\mconc}}^\slagr - \mconc} \in \bigo{\sqrt{\frac{(\card{E} + 1)\log (p)}{N}}} \text{ as }N\tendsto\infty.
\end{equation*}
{Convergence rates in other norms are provided in
\citet{ravikumar2011}, and they have been generalized by
\citet{lam2009} for other penalty functions.}

\section{Bayesian {model selection} and estimation}\label{sec:bayes}
{Consider a continuous multivariate family $\fparam{\vparam}$
parametrized by $\vparam$, and denote as $f(\msample \sgiven \vparam)$ the
density function of a random sample $\msrand$ from $P \in \fparam{\vparam}$ for
a given value of $\vparam$. In Bayesian statistics, $\vparam$ is treated as a
random variable with known distribution, $f(\vparam)$, usually called the
\emph{prior} distribution of $\vparam$. Inference is then performed based on the
value of $f(\vparam \sgiven \msample) \propto f(\vparam)f(\msample
\sgiven \vparam)$, the \emph{posterior} distribution of $\vparam$ given the
information in $\msrand = \msample$. 

In Gaussian Markov models, $\vparam = (\vmean, \mconc, \graph{G})$, where in our
case $\graph{G}$ is either undirected or acyclic directed.} Therefore the target
probability is $f(\graph{G}, \vmean, \mconc \sgiven \msample) \propto f(\msample
\sgiven \vmean, \mconc, \graph{G}) \, f(\vmean, \mconc \sgiven \graph{G})\,
f(\graph{G})$. Integrating out $\vmean$ and $\mconc$, we obtain the posterior
density of model $\graph{G}$, $f(\graph{G} \sgiven \msample)\propto f(\msample
\sgiven \graph{G})\,f(\graph{G})$. The prior for the graph space,
$f(\graph{G})$, is usually set as uniform. However, this choice is biased
towards middle size graphs, and thus other prior distributions
\citep[e.g.]{scutari2013} have been proposed{; see
\citet{massam2018} \S10.4.1 and references therein for a recent detailed
overview.  Bayesian inference for Gaussian graphical models, is usually meant
for moderate sample sizes, since it relies on sampling from the resulting
posterior distribution, which becomes infeasible in high dimensions (see e.g.
\citealp{jones2005} or \citealp{massam2018}).}

In the following, the \textit{$p$-variate Wishart distribution} will be denoted
as $\dmwis{p}{n}{\mat{\Lambda}}$ with $n \in \sreal$, $n > p -1$ and
$\mat{\Lambda} \in \mgens{p}{p}{\sreal}$, $\mat{\Lambda} \succ 0$; analogously,
the \textit{$p$-variate inverse Wishart distribution} will be
$\dmiwis{p}{\nu}{\mat{\Psi}}$ with $\nu \in \sreal$, $\nu > p -1$ and
$\mat{\Psi} \in \mgens{p}{p}{\sreal}$, $\mat{\Psi} \succ 0$. 

\subsection{Hyper Markov laws}

{When $\graph{G}$ is undirected and decomposable, and assuming $\vmean = \vecb{0}$,
\citet{dawid1993} proposed for $f(\mconc \sgiven \graph{G})$ what are
known as the \textit{hyper Markov laws}. {These
are defined in terms of properties of the graph associated with the Markov
model, mimicking Markov properties. Specifically,} let $\vrandparam$ be a
random variable taking values over $\ggmm{\graph{G}}$ {and for
subsets $A, B \subseteq V$, denote as $\vparam_A$ and $\vparam_{B \sgiven A}$
the parameters of the marginal distribution of $\vrand_A$ and the conditional
distribution of $\vrand_A$ given values of $\vrand_B$, respectively}. The probability
distribution of $\vrandparam$ is said to be \textit{(weakly) hyper $\graph{G}$ -
Markov} if, for {any} decomposition $(A,B{, S})$ of $\graph{G}$, it holds that
$\ci{\vrandparam_{A\cup S}}{\vrandparam_{B\cup S}}{\vrandparam_{{S}}}$; if it further holds
${\vrandparam_{B \cup S\sgiven A \cup S}}\sindep \vrandparam_{A \cup S}$, it is called \textit{strongly
hyper {$\graph{G}$-}Markov}. {For chordal graphs,} if the probability
distribution of $\vrandparam$ is strongly hyper {$\graph{G}$-}Markov with
respect to $\graph{G}$, then the probability distribution of
$\vrandparam\sgiven\msample$ is the unique (strong) hyper {$\graph{G}$-}Markov
distribution specified by the clique-marginal distributions
$\{{\fprobm{\vrandparam_C \sgiven \msample_{C}}}\st C\in\scliques{\graph{G}}\}${;
and,} when densities exist, $f(\vrandparam_C\sgiven\msample)\propto
f(\msample_{C}\sgiven\vrandparam_C)f(\vrandparam_C)$
\citep{dawid1993}{, where $\msample_C$ stands for all the
observations in $\msample$ corresponding to the variables in $C$}. That is, under these
assumptions, it is possible to localize computations over the graph cliques when
performing Bayesian inference. 

{{In a multivariate Gaussian distribution $\mathcal{N}_p(\vecb{0},
\mcov$)}, the inverse Wishart is a conjugate
prior for $\mcov$; that is,} if $\mcov \sdist
\dmiwis{p}{{\nu}}{{\mat{\Psi}}}$, then $\mcov \sgiven
{\msprod/N} \sdist \dmiwis{p}{N + {\nu}}{\msprod +
{\mat{\Psi}}}$ {(recall $\msprod = \msample\msample^t$)}. We
can thus construct the \textit{hyper inverse Wishart
distribution}, as the unique hyper Markov distribution
associated with inverse Wishart clique marginals{:}
$\mcov_{CC} \sdist  \dmiwis{\card{C}}{\nu}{\mat{\Psi}{^C}}$,
for each clique $C\in\scliques{\graph{G}}$. This hyper Markov
distribution is denoted as $\dmhiwis{p}{{\nu}}{\mat{\Psi}}$,
where $\mat{\Psi}\in\msympdefs$ such that $\mat{\Psi}_{CC} =
\mat{\Psi}^C$ for each clique $C\in\scliques{\graph{G}}$. {From the
discussion above, we know that} this
distribution is strongly hyper {$\graph{G}$-}Markov. The main
advantage of this prior is that it has many properties that
mirror those for Markov models, since hyper Markov
distributions are also defined in terms of an underlying
graph.

{Since its introduction, the hyper inverse Wishart prior for
decomposable graphs has been extensively studied.} The explicit expression for its
density is devised in, e.g., \cite{giudici1996} or
\citet{roverato2000}. In order to set its parameters, a hierarchical 
approach such as in \cite{giudici1999} can be followed, where $\nu$ and
$\mat{\Psi}$ are assumed to have a Gamma and Wishart distribution, respectively. {Since the absent edges of $\graph{G}$
correspond to zeros in $\mconc$ (Equation \eqref{eq:ug:mvn:indep}),
\citet{roverato2000} derives the distribution induced on $\mconc$ by assuming
$f(\mcov \sgiven \graph{G}) = \dmhiwis{p}{{\nu}}{\mat{\Psi}}$. He shows that in that case,
the density is proportional to that of a Wishart matrix conditioned on the event
$\mconc \in \mugmm{\graph{G}}$, and calls such prior distribution on $\mconc$
the \textit{$\graph{G}$-conditional Wishart}. 
Recently, \citet{massam2018} \S10.3.2 has provided a detailed
overview of the properties of the hyper inverse Wishart, and technical
considerations as how to sample from it or perform Bayesian model selection
using Bayes factors. \citet{letac2007} generalize both the
$\graph{G}$-conditional and hyper inverse Wishart to a broader conjugate family,
allowing for more than one shape parameter, which is used for model selection by
\citet{rajaratnam2008} \citep[see
also][\S10.3.3]{massam2018}.}

The hyper inverse Wishart has been extended to non-chordal graphs by
\citet{roverato2002}, based on properties of the Isserlis matrix of $\mcov$
\citep{roverato1998}. { However, Bayesian model selection in this
scenario requires the evaluation of the $\graph{G}$-conditional
Wishart normalizing constant, which becomes a problem since it did not have a known closed-form
expression for a general non-chordal $\graph{G}$ until very recently \citep{uhler2018}. Much of the literature
therefore has been devoted to this issue: \citet{atay-kayis2005} analysed the
Cholesky decomposition of $\mconc$ and its relation with the cone
$\mugmm{\graph{G}}$ and positive definite matrix completions (Section
\ref{sec:mle}), \citet{carvalho2007} and \citet{wang2010} used such theoretical
analysis to provide a direct sampler from the hyper inverse Wishart prior, etc.
A recent detailed presentation of this computational body of research can be
found in \citet{massam2018}, \S10.4. Note that, although \citet{uhler2018}
provide exact formulas and examples for special types of graphs, it still
remains to find efficient methods for their computation.
}

\subsection{Priors for acyclic directed models}
The methodology by \citet{geiger2002} for acyclic directed Gaussian
Markov models can be seen as an extension of hyper Markov distributions to such
context, since they both coincide for chordal skeletons.

If $\graph{D}_c$ is an arbitrary
complete digraph, then{, under some assumptions on
$f(\vmean, \mconc \sgiven \graph{D})$
and $f(\msample \sgiven \graph{D}, \vmean, \mconc)$, computations can be localized as}
\begin{equation}\label{eq:marglike}
	f(\msample \sgiven \graph{D}) = \prod_{v \in V}\frac{f(\msample_{\set{v}
	\cup \pa_{\graph{D}}(v)} \sgiven
	\graph{D}_c)}{f(\msample_{\pa_{\graph{D}}(v)} \sgiven \graph{D}_c)},
\end{equation}
The posterior in Equation \eqref{eq:marglike} is equal among Markov
equivalent acyclic digraphs \citep{geiger2002}.

The conjugate prior for
$(\vmean, \mconc)$ is {the} normal-Wishart distribution, where $\mconc
\sdist \dmwis{p}{\alpha_{\mconc}}{{\mat{\Lambda}}}$ and {$\vmean \sgiven \mconc
 \sdist \dmnorm{p}{{\vmean_0}}{(\alpha_{\vmean}\mconc)^{-1}}$}.
 {This yields a normal-Wishart posterior distribution for $(\vmean,
 \mconc) \sgiven \graph{D}_c$. Using this,} 
\cite{geiger1994}, obtain {an explicit expression for each factor in Equation
\eqref{eq:marglike}}: for $U\subseteq V$,
\begin{equation*}\label{eq:ghright}
	f(\msample_U \sgiven\graph{D}_c) = \left(\frac{\alpha_{\vmean}}{\alpha_{\vmean} + N}\right)^{\frac{{\card{U}}}{2}}2\pi^{-\frac{lN}{2}}\frac{\Gamma{_{\card{U}}}\left(\frac{N+\alpha_{\mconc} - p + {\card{U}}}{2}\right)}{\Gamma{_{\card{U}}}\left(\frac{\alpha_{\mconc} - p + {\card{U}}}{2}\right)}\frac{\abs{{\mat{\Lambda}}_{{UU}}}^{\frac{\alpha_{\mconc} - p + {\card{U}}}{2}}}{\abs{\mat{R}_{UU}}^{\frac{\alpha_{\mconc} - p + {\card{U}} + N}{2}}},
\end{equation*}
where {$\Gamma_p(\cdot)$ is the $p$-dimensional Gamma function, and}
\[
	\mat{R} = {\mat{\Lambda}} + \msprod + \frac{N\alpha_{\mconc}}{N +
	\alpha_{\mconc}}({\vmean_0} - \vsmean)({\vmean_0} - \vsmean)^t.
\]
Furthermore, \citet{geiger2002} characterize the normal-Wishart prior for $(\vmean,
\mconc)$ {as} the only distribution satisf{ying the global
parameter independence} assumption,
\begin{equation*}
	f(\vrandparam \sgiven \graph{D}) =
	\prod_{v \in V}f(\sparam_v \sgiven \graph{D})
\end{equation*}
for every $P_{\vecb{X}} \in \ggmm{\graph{D}}$. {This condition is
required for Equation \eqref{eq:marglike} to hold. 

The above priors have been used by \citet{consonni2012} and \citet{altomare2013}
for objective Bayesian model selection, where $f(\vparam \sgiven
\graph{D})$ might be improper. For overcoming this, they use fractional Bayes
factors \citep{ohagan1995}, which had been also previously used by
\citet{carvalho2009} for chordal undirected models \citep[see][\S10.6 and
references therein for more details]{massam2018}. Recently, \citet{bendavid2016}
have proposed a family of priors extending those by \citet{geiger2002} but including more
shape parameters, that is, mimicking those in \citet{letac2007} for undirected
models. See \citet{rajaratnam2012} and \citet{cao2019} for further discussion on
these priors.}

\section{Higher level Markov model classes with mixed graphs}\label{sec:higher}
{
As we have seen throughout this review, the classes of acyclic directed and
undirected Markov models are intimately related. Therefore, one approach for
their unified treatment could be to step to a higher level, and define new
Markov model classes containing them as subclasses. In this section we will
overview this approach, which has been particularly active in the past few
years. The graphs used for these new Markov models are usually called mixed
graphs, because unlike purely undirected or acyclic directed graphs, they allow
for more than one edge type. We do not aim in this section for a thorough
account of the achievements and drawbacks of the different developments since
that would take another full paper.

Chain graphs are the first higher level attempt at this unification: they allow
two edge types, directed and undirected, and forbid semi-directed cycles.  \citet{drton2009} provides a
unifying view of these model classes, focusing on their discrete parametrization: both the
undirected and directed edges can have two different interpretations, thus
giving rise to four different chain graph model classes.  Among them, \textit{AMP chain
graphs} \citep{andersson2001}, and \textit{LWF chain graphs}
\citep{lauritzen1989,frydenberg1990}, named in such way because of the
respective paper authors, contain both acyclic directed and undirected Markov
model classes.

\textit{Multivariate regression} (MVR) \textit{chain graphs} \citep{cox1993,cox1996}, or Type IV in
\citet{drton2009}, are possibly the most traditional ones and can be viewed as
a special case of the path diagrams by \citet{wright1934}. Although they do not
contain undirected models, their extension, \textit{regression graphs}
\citep{wermuth2012,wermuth2011}, do contain both classes treated in this review,
by allowing up to three edge types. 
The class of regression graphs allows to represent additional
relationships in classical sequences of multivariate regressions by means of a
bi-directed edge, which could not be otherwise modelled using only the acyclic
directed or undirected graphs. These bi-directed edges represent interactions
with \emph{latent variables}. The recovery of latent variables is specially
relevant in social studies, where the presence of these \emph{confounding}
variables may affect the prediction. It
has been recently shown that a Gaussian MVR chain graph is Markov equivalent to
an acyclic directed Gaussian Markov model with latent variables when its
bidirected part is chordal \citep{fox2015}. Pure Gaussian bidirected graphs represent
marginal independences among variables, therefore they impose zero constraints
directly in the covariance $\mat{\Sigma}$. 

Finally, the \textit{Type III chain graph} so far has not
been devoted much attention \citep{lauritzen2018}. All of the mentioned chain graph
model classes are smooth, whereas for their discrete counterparts only the
classes of LWF and
multivariate regression chain graphs consist of smooth models
\citep{drton2009}.

The semi-directed cycle constraint on chain graphs can be relaxed, and new
graphical model classes are obtained by forbidding only directed cycles. By
doing so, we arrive at three different classes of what are called acyclic
directed mixed graphs: the so-called \textit{original acyclic directed mixed
graph}
\citep[oADMG,][]{richardson2003}, the \textit{alternative} \citep[aADMG,][]{pena2016}, and
\textit{UDAGs}
\citep{pena2018}, which relax MVR, AMP and LWF chain graphs, respectively.
Each oADMG model contains a model obtained from a
Bayesian network after marginalizing some of its nodes, (the latent variables).
However, other constraints may arise after marginalizing that cannot be
represented in terms of conditional independence with this class, for
example, the \emph{Verma constraints} \citep[\S7.3.1]{richardson2002} and
\emph{inequality constraints} \citep{drton2012}. In order to deal with these,
\citet{richardson2012} introduced nested Markov models which also allow for
hyper-edges between more than two nodes; however, we are not aware of any
Gaussian parametrization.

The classes of both oADMGs and aADMGs are subsumed by the class of \textit{ADM
graphs} \citep{pena2018}, consisting, naturally, of three edge types. When
parametrized with the Gaussian distribution, ADM graph models can be represented
as recursive linear equations with two blocks of variables and possibly
correlated errors \citep{koster1999,spirtes1995,pena2016,pena2016b}. Bidirected
edges in these models represent latent confounding effects,
whereas undirected edges account for dependence between the errors. Note that,
although the classes of ADM graphs and regression graphs allow the same edge
types, they are not equivalent since the former contains AMP chain graph models,
while the latter doesn't.

There are other models allowing for up to three edge types, besides the already
mentioned ADM and regression graphs: \textit{anterial} and \textit{chain mixed
graphs} \citep{sadeghi2016}, \textit{ribbonless graphs} \citep{sadeghi2013},
\textit{MC graphs} \citep{koster2002}, \textit{summary graphs}
\citep{wermuth2011,cox1996}, \textit{ancestral graphs} \citep{richardson2002},
etc. These model classes share rich relationships, which have been recently
discussed by \citet{lauritzen2018}. Ancestral graphs extend regression graphs
by relaxing the cycle constraint, but  they are not a maximal class; that is, if
an edge is removed from the graph, we may remain on the same Markov model.
Maximality is convenient because it is what allows to define pairwise Markov
properties, so that each edge absent implies a conditional independence.
Fortunately, for an arbitrary ancestral graph we may always find a maximal one
which is Markov equivalent to it, therefore many times authors speak of the
class of \textit{maximal ancestral graphs} (MAGs). This class is {closed} under
marginalization and conditioning, and every MAG can be obtained from an acyclic
digraph after performing such operations on its nodes. Just as marginalization
leads to latent confounders, conditioning is sometimes called \emph{selection
bias} in the literature on social sciences.

The class of summary graphs, although in correspondence with ancestral graphs,
is not easily parameterized. One of the main drawbacks is that they allow more
than one edge type between the nodes, which means that in principle more than
one parameter can be associated between a pair of variables. Furthermore, they
are not maximal and thus cannot have a pairwise Markov property, which implies
that fewer independences can be deduced from the model. Most of the other
three-edge-type models mentioned share these drawbacks for defining a
parametrization \citep{richardson2002,sadeghi2012b}.

The proliferation of higher level Markov model classes has led
\citet{lauritzen2018} to recently propose a class of mixed graphs consisting of
up to four edge types, in an attempt to unify most of them under a unique Markov
property \citep[see also][]{sadeghi2014,evans2018}. Nowadays, a great amount of research is
focused on characterizing basic foundational properties for these higher level
models: for example, Markov equivalence, definition and equivalence of Markov
properties, factorization properties, etc.
}

\section{Relaxing the Gaussian assumption}\label{sec:alt}
In some real problems, the Gaussian assumption is too restrictive, and thus some
alternative models to overcome this have been proposed. Although these are
outside the scope of this review, we will survey here the main proposals to
relax the Gaussian assumption.

As we have seen, Gaussian
Bayesian networks are equivalent to a set of recursive regressions
where the errors are Gaussian. In \citet{shimizu2006}, an analogous model is
proposed, called LiNGAM, where the errors are assumed to be non Gaussian. The work by
\citet{loh2014} generalizes further generalizes this by not making
distributional assumptions on the errors.
As an
alternative, \citet{peters2014b} and \citet{buhlmann2014} maintain Gaussian
errors but the additive regression
is now assumed to be non linear. Other families of continuous distributions that have been used for
parametrizing Markov and Bayesian networks are nonparametric Gaussian copulas
\citep{liu2009}
and elliptical distributions \citep{vogel2011}, both of which generalize the Gaussian distribution.
Copula graphical models are usually referred to as `nonparanormal' models, and
model selection and estimation have been researched by
\citet{harris2013}, \citet{xue2012} and
\citet{liu2012}, including high-dimensional scenarios. Estimation results for elliptical graphical models
have been obtained by \citet{vogel2014}.

Another approach is to extend the model and allow for both discrete and continuous variables. In such
case, a challenge is posed specially on inference in Bayesian networks, where the usual operations may not allow for
a direct and efficient implementation as in the pure cases. The main source of
this problem is the integration that appears in marginalization of continuous
variables.
To overcome this, the classical approach is to use the conditional Gaussian
distribution \citep{olkin1961}.
It is characterized by a multinomial distribution on the discrete variables and
a Gaussian distribution for the continuous variables when conditioned on the
discrete ones. Therefore, it contains the pure multinomial and Gaussian models
as particular cases. Markov properties of this distribution with respect to an undirected graph
were defined by \citet{lauritzen1989}. With respect to an acyclic digraph, a
further assumption is that no discrete variable may have continuous parents,
which leads to conditional linear Gaussian Bayesian networks
\citep{lauritzen1992}. Exact inference in these networks is applicable thanks to
these constraints imposed on the network topology.

In order to avoid the structural constraints of conditional linear Gaussian
Bayesian networks, nonparametric density estimation techniques have been
proposed.
\citet{moral2001} approximated the joint density by mixtures of
truncated exponentials. In this model, discrete nodes with continuous parents are allowed, while exact
inference remains possible. A similar approach is that of \citet{shenoy2011},
where mixtures of polynomials are used instead for approximating the joint
density. These two models have been generalized by \citet{langseth2012} as mixtures
of truncated basis functions.
However, there are limited results about maximum likelihood estimation and model
selection for these models \citep{langseth2010,langseth2014,varando2015}.

\section{Main application areas}\label{sec:app}
Graphical or Markov models have been widely applied since their conception and
continue to be nowadays an essential tool in many fields, since they are intuitive
for visualizing the associations between the components in a system. We will first outline
applications of Gaussian Markov and Gaussian Bayesian networks and then
illustrate other areas where graphical models have played an important role.

Markov and Bayesian networks with Gaussian parametrization have been specially
useful in biomedical sciences. For
example, Gaussian Bayesian networks have been used for extracting
knowledge from fMRI studies \citep{mumford2014,zhou2016}, where nodes are identified with brain
regions, and arrows are interpreted as direct influences between the respective
regions. Another example where both models have been applied is the modelling of gene regulatory
networks, which are high-dimensional and complex by nature. In fact,
the challenge posed by this problem has served as an impulse for methodological developments in
both models. A vast amount of
literature can be found regarding the main computational
aspects
involved on this subject, as well as interpretability issues, see
\citet{lauritzen2003}, \citet{friedman2004}, \citet{markowetz2007} and
\citet{ness2016} for reviews.

In social sciences, Bayesian networks have been used since their conception, in
fact, we could say that research in this application area helped to settle the
foundations of graphical models \citep{kiiveri1982}. In terms of
interpretability, the directed arcs in Bayesian networks are usually given a causal
interpretation \citep{pearl2000,cox1996}, since ultimately the main goal of
social studies is to identify the causes of a resulting event of interest.
In recent years, graphical models have found a natural area of application which
is social network analysis \citep{farasat2015}, which include problems such as
influence analysis, privacy protection, web browsing, etc.

Some other traditional models from bioinformatics can also be seen as graphical models.
These include phylogenetic trees, which model evolutionary relationships between
different species or organisms, and pedigrees, which are diagrams showing the
occurrence and variants of a gene from one generation of organisms to the next
\citep{jordan2004}. Apart from fMRI studies, Bayesian networks have also been
applied in different subareas of neuroscience \citep{bielza2014}, such as
morphological and electrophysiological studies. Finally, other, more technical
application areas include information
retrieval \citep{decampos2004}, where relevant documents about some matter are collected from an
available set of sources, and linguistics, with subfields such as speech recognition \citep{deng2013},
and natural language processing \citep{cambria2014}.

\section*{Acknowledgements}
This work has been supported by the Spanish Ministry of Science, Innovation and
Universities through the predoctoral grant FPU15/03797 holded by Irene Córdoba,
and through the TIN2016-79684-P project.
\bibliographystyle{elsarticle-harv}
\bibliography{revision_GMM}

\begin{thebibliography}{172}
\expandafter\ifx\csname natexlab\endcsname\relax\def\natexlab#1{#1}\fi
\expandafter\ifx\csname url\endcsname\relax
  \def\url#1{\texttt{#1}}\fi
\expandafter\ifx\csname urlprefix\endcsname\relax\def\urlprefix{URL }\fi

\bibitem[{Altomare et~al.(2013)Altomare, Consonni, and La~Rocca}]{altomare2013}
Altomare, D., Consonni, G., La~Rocca, L., 2013. Objective {B}ayesian search of
  {G}aussian directed acyclic graphical models for ordered variables with
  non-local priors. Biometrics 69~(2), 478--487.

\bibitem[{Anderson(1973)}]{anderson1973}
Anderson, T.~W., 1973. Asymptotically efficient estimation of covariance
  matrices with linear structure. Ann. Stat. 1~(1), 135--141.

\bibitem[{Anderson(2003)}]{anderson2003}
Anderson, T.~W., 2003. An Introduction to Multivariate Statistical Analysis,
  3rd Edition. John Wiley \& Sons.

\bibitem[{Andersson et~al.(2001)Andersson, Madigan, and
  Perlman}]{andersson2001}
Andersson, S.~A., Madigan, D., Perlman, M.~D., 2001. Alternative {M}arkov
  properties for chain graphs. Scand. J. Stat. 28~(1), 33--85.

\bibitem[{Andersson and Perlman(1998)}]{andersson1998}
Andersson, S.~A., Perlman, M.~D., 1998. Normal linear regression models with
  recursive graphical {M}arkov structure. J. Multivar. Anal. 66~(2), 133 --
  187.

\bibitem[{{Aragam} et~al.(2017){Aragam}, {Amini}, and {Zhou}}]{aragam2017}
{Aragam}, B., {Amini}, A.~A., {Zhou}, Q., 2017. Learning directed acyclic
  graphs with penalized neighbourhood regression. arXiv:1511.08963.

\bibitem[{Aragam and Zhou(2015)}]{aragam2015}
Aragam, B., Zhou, Q., 2015. Concave penalized estimation of sparse {G}aussian
  {B}ayesian networks. J. Mach. Learn. Res. 16, 2273--2328.

\bibitem[{Atay-Kayis and Massam(2005)}]{atay-kayis2005}
Atay-Kayis, A., Massam, H., 2005. A {M}onte {C}arlo method for computing the
  marginal likelihood in nondecomposable {G}aussian graphical models.
  Biometrika 92~(2), 317--335.

\bibitem[{Banerjee et~al.(2008)Banerjee, El~Ghaoui, and
  d'Aspremont}]{banerjee2008}
Banerjee, O., El~Ghaoui, L., d'Aspremont, A., 2008. Model selection through
  sparse maximum likelihood estimation for multivariate {G}aussian or binary
  data. J. Mach. Learn. Res. 9, 485--516.

\bibitem[{Barndorff-Nielsen(1978)}]{barndorff-nielsen1978}
Barndorff-Nielsen, O., 1978. Information and Exponential Families in
  Statistical Theory. John Wiley \& Sons.

\bibitem[{{Ben-David} et~al.(2016){Ben-David}, {Li}, {Massam}, and
  {Rajaratnam}}]{bendavid2016}
{Ben-David}, E., {Li}, T., {Massam}, H., {Rajaratnam}, B., 2016. {High
  dimensional {B}ayesian inference for {G}aussian directed acyclic graph
  models}. arXiv:1109.4371.

\bibitem[{Ben-David and Rajaratnam(2012)}]{bendavid2012}
Ben-David, E., Rajaratnam, B., 2012. Positive definite completion problems for
  {B}ayesian networks. SIAM J. Matrix Anal. Appl. 33~(2), 617--638.

\bibitem[{Besag(1974)}]{besag1974}
Besag, J., 1974. Spatial interaction and the statistical analysis of lattice
  systems. J. R. Stat. Soc. Ser. B Stat. Methodol. 36~(2), 192--236.

\bibitem[{Bielza and Larra\~naga(2014)}]{bielza2014}
Bielza, C., Larra\~naga, P., 2014. Bayesian networks in neuroscience: {A}
  survey. Front. Comput. Neurosci. 8, 131.

\bibitem[{B\"uhlmann et~al.(2014)B\"uhlmann, Peters, and Ernest}]{buhlmann2014}
B\"uhlmann, P., Peters, J., Ernest, J., 2014. {CAM}: {C}ausal additive models,
  high-dimensional order search and penalized regression. Ann. Stat. 42~(6),
  2526--2556.

\bibitem[{B\"uhlmann and van~de Geer(2011)}]{buehlmann2011}
B\"uhlmann, P., van~de Geer, S., 2011. Statistics for High-Dimensional Data:
  Methods, Theory and Applications. Springer.

\bibitem[{Cambria and White(2014)}]{cambria2014}
Cambria, E., White, B., 2014. Jumping {NLP} curves: {A} review of natural
  language processing research. IEEE Comput. Intell. Mag. 9~(2), 48--57.

\bibitem[{Cao et~al.(2019)Cao, Khare, and Ghosh}]{cao2019}
Cao, X., Khare, K., Ghosh, M., 2019. Posterior graph selection and estimation
  consistency for high-dimensional {B}ayesian {DAG} models. Ann. Stat. 47~(1),
  319--348.

\bibitem[{Carvalho et~al.(2007)Carvalho, Massam, and West}]{carvalho2007}
Carvalho, C.~M., Massam, H., West, M., 2007. Simulation of hyper-inverse
  {W}ishart distributions in graphical models. Biometrika 94~(3), 647--659.

\bibitem[{Carvalho and Scott(2009)}]{carvalho2009}
Carvalho, C.~M., Scott, J.~G., 2009. Objective {B}ayesian model selection in
  {G}aussian graphical models. Biometrika 96~(3), 497--512.

\bibitem[{Castelo and Roverato(2006)}]{castelo2006}
Castelo, R., Roverato, A., 2006. A robust procedure for {G}aussian graphical
  model search from microarray data with {\it p} larger than {\it n}. J. Mach.
  Learn. Res. 6, 2621--2650.

\bibitem[{Colombo and Maathuis(2014)}]{colombo2014}
Colombo, D., Maathuis, M.~H., 2014. Order-independent constraint-based causal
  structure learning. J. Mach. Learn. Res. 15, 3921--3962.

\bibitem[{Consonni and Rocca(2012)}]{consonni2012}
Consonni, G., Rocca, L.~L., 2012. Objective {B}ayes factors for {G}aussian
  directed acyclic graphical models. Scand. J. Stat. 39~(4), 743--756.

\bibitem[{Cox and Wermuth(1993)}]{cox1993}
Cox, D.~R., Wermuth, N., 1993. Linear dependencies represented by chain graphs.
  Stat. Sci. 8~(3), 204--218.

\bibitem[{Cox and Wermuth(1996)}]{cox1996}
Cox, D.~R., Wermuth, N., 1996. Multivariate dependencies: {M}odels, analysis
  and interpretation. Chapman \& Hall.

\bibitem[{Daly et~al.(2011)Daly, Shen, and Aitken}]{daly2011}
Daly, R., Shen, Q., Aitken, S., 2011. Learning {B}ayesian networks:
  {A}pproaches and issues. Knowl. Eng. Rev. 26, 99--157.

\bibitem[{Darroch et~al.(1980)Darroch, Lauritzen, and Speed}]{darroch1980}
Darroch, J.~N., Lauritzen, S.~L., Speed, T.~P., 1980. Markov fields and
  log-linear interaction models for contingency tables. Ann. Stat. 8~(3),
  522--539.

\bibitem[{Dawid(1979)}]{dawid1979}
Dawid, A.~P., 1979. Conditional independence in statistical theory. J. R. Stat.
  Soc. Ser. B Stat. Methodol. 41~(1), 1--31.

\bibitem[{Dawid(1980)}]{dawid1980}
Dawid, A.~P., 1980. Conditional independence for statistical operations. Ann.
  Stat. 8~(3), 598--617.

\bibitem[{Dawid(2001)}]{dawid2001}
Dawid, A.~P., 2001. Separoids: A mathematical framework for conditional
  independence and irrelevance. Ann. Math. Artif. Intell. 32~(1), 335--372.

\bibitem[{Dawid and Lauritzen(1993)}]{dawid1993}
Dawid, A.~P., Lauritzen, S.~L., 1993. Hyper {Markov} laws in the statistical
  analysis of decomposable graphical models. Ann. Stat. 21~(3), 1272--1317.

\bibitem[{de~Campos et~al.(2004)de~Campos, Fernández-Luna, and
  Huete}]{decampos2004}
de~Campos, L.~M., Fernández-Luna, J.~M., Huete, J.~F., 2004. Bayesian networks
  and information retrieval: an introduction to the special issue. Inf.
  Process. Manag. 40~(5), 727 -- 733.

\bibitem[{de~la Fuente et~al.(2004)de~la Fuente, Bing, Hoeschele, and
  Mendes}]{fuente2004}
de~la Fuente, A., Bing, N., Hoeschele, I., Mendes, P., 2004. Discovery of
  meaningful associations in genomic data using partial correlation
  coefficients. Bioinformatics 20~(18), 3565--3574.

\bibitem[{Dempster(1972)}]{dempster1972}
Dempster, A.~P., 1972. Covariance selection. Biometrics 28~(1), 157--175.

\bibitem[{Deng and Li(2013)}]{deng2013}
Deng, L., Li, X., 2013. Machine learning paradigms for speech recognition: An
  overview. IEEE Trans. Audio, Speech, Lang. Process. 21~(5), 1060--1089.

\bibitem[{Drton(2009)}]{drton2009}
Drton, M., 2009. Discrete chain graph models. Bernoulli 15~(3), 736--753.

\bibitem[{Drton et~al.(2012)Drton, Fox, and K{\"a}ufl}]{drton2012}
Drton, M., Fox, C., K{\"a}ufl, A., Jun 2012. Comments on: Sequences of
  regressions and their independencies. TEST 21~(2), 255--261.

\bibitem[{Drton and Perlman(2004)}]{drton2004}
Drton, M., Perlman, M.~D., 2004. Model selection for {G}aussian concentration
  graphs. Biometrika 91~(3), 591--602.

\bibitem[{Drton and Perlman(2007)}]{drton2007}
Drton, M., Perlman, M.~D., 2007. Multiple testing and error control in
  {G}aussian graphical model selection. Stat. Sci. 22~(3), 430--449.

\bibitem[{Drton and Perlman(2008)}]{drton2008}
Drton, M., Perlman, M.~D., 2008. A {SIN}ful approach to {G}aussian graphical
  model selection. J. Stat. Plan. Inference 138~(4), 1179--1200.

\bibitem[{Eriksen(1996)}]{eriksen1996}
Eriksen, P.~S., 1996. Tests in covariance selection models. Scand. J. Stat.
  23~(3), 275--284.

\bibitem[{Evans(2018)}]{evans2018}
Evans, R., 2018. Markov properties for mixed graphical models. In: Handbook of
  Graphical Models. CRC Press, pp. 57--78.

\bibitem[{Farasat et~al.(2015)Farasat, Nikolaev, Srihari, and
  Blair}]{farasat2015}
Farasat, A., Nikolaev, A., Srihari, S.~N., Blair, R.~H., 2015. Probabilistic
  graphical models in modern social network analysis. Soc. Netw. Anal. Min.
  5~(1), 62.

\bibitem[{Fox et~al.(2015)Fox, Käufl, and Drton}]{fox2015}
Fox, C.~J., Käufl, A., Drton, M., 2015. On the causal interpretation of
  acyclic mixed graphs under multivariate normality. Linear Algebra Its Appl.
  473~(Suppl. C), 93--113.

\bibitem[{Friedman et~al.(2008)Friedman, Hastie, and Tibshirani}]{friedman2008}
Friedman, J., Hastie, T., Tibshirani, R., 2008. Sparse inverse covariance
  estimation with the graphical lasso. Biostatistics 9~(3), 432--441.

\bibitem[{Friedman(2004)}]{friedman2004}
Friedman, N., 2004. Inferring cellular networks using probabilistic graphical
  models. Science 303~(5659), 799--805.

\bibitem[{Frydenberg(1990)}]{frydenberg1990}
Frydenberg, M., 1990. The chain graph {M}arkov property. Scand. J. Stat.
  17~(4), 333--353.

\bibitem[{Frydenberg and Lauritzen(1989)}]{frydenberg1989}
Frydenberg, M., Lauritzen, S., 1989. Decomposition of maximum likelihood in
  mixed graphical interaction models. Biometrika 76~(3), 539--555.

\bibitem[{Geiger and Heckerman(1994)}]{geiger1994}
Geiger, D., Heckerman, D., 1994. Learning {G}aussian networks. In: Proc. of the
  Tenth Conference on Uncertainty in Artificial Intelligence. Morgan Kaufmann,
  San Francisco, pp. 235--243.

\bibitem[{Geiger and Heckerman(2002)}]{geiger2002}
Geiger, D., Heckerman, D., 2002. Parameter priors for directed acyclic
  graphical models and the characterization of several probability
  distributions. Ann. Stat. 30~(5), 1412--1440.

\bibitem[{Geiger and Pearl(1990)}]{geiger1990b}
Geiger, D., Pearl, J., 1990. On the logic of causal models. In: Proc. of the
  Fourth Annual Conference on Uncertainty in Artificial Intelligence. AUAI
  Press, Corvallis, pp. 3--14.

\bibitem[{Geiger and Pearl(1993)}]{geiger1993}
Geiger, D., Pearl, J., 1993. Logical and algorithmic properties of conditional
  independence and graphical models. Ann. Stat. 21~(4), 2001--2021.

\bibitem[{Gillispie and Perlman(2002)}]{gillispie2002}
Gillispie, S.~B., Perlman, M.~D., 2002. The size distribution for {M}arkov
  equivalence classes of acyclic digraph models. Artif. Intell. 141~(1–2),
  137 -- 155.

\bibitem[{Giudici(1996)}]{giudici1996}
Giudici, P., 1996. Learning in graphical {G}aussian models. Bayesian Stat. 5,
  621--628.

\bibitem[{Giudici and Green(1999)}]{giudici1999}
Giudici, P., Green, P.~J., 1999. {Decomposable graphical {G}aussian model
  determination}. Biometrika 86~(4), 785--801.

\bibitem[{Gr{\"a}del and V{\"a}{\"a}n{\"a}nen(2013)}]{graedel2013}
Gr{\"a}del, E., V{\"a}{\"a}n{\"a}nen, J., 2013. Dependence and independence.
  Stud. Log. 101~(2), 399--410.

\bibitem[{Grimmett(1973)}]{grimmett1973}
Grimmett, G.~R., 1973. A theorem about random fields. Bull. Lond. Math. Soc.
  5~(1), 81--84.

\bibitem[{Grone et~al.(1984)Grone, Johnson, Sá, and Wolkowicz}]{grone1984}
Grone, R., Johnson, C.~R., Sá, E.~M., Wolkowicz, H., 1984. Positive definite
  completions of partial hermitian matrices. Linear Algebra Its Appl. 58, 109
  -- 124.

\bibitem[{Hammersley and Clifford(1971)}]{hammersley1971}
Hammersley, J.~M., Clifford, P., 1971. Markov fields on finite graphs and
  lattices, {U}npublished manuscript.

\bibitem[{Harris and Drton(2013)}]{harris2013}
Harris, N., Drton, M., 2013. {PC} algorithm for nonparanormal graphical models.
  J. Mach. Learn. Res. 14, 3365--3383.

\bibitem[{He et~al.(2015)He, Jia, and Yu}]{he2016b}
He, Y., Jia, J., Yu, B., 2015. Counting and exploring sizes of {M}arkov
  equivalence classes of directed acyclic graphs. J. Mach. Learn. Res. 16,
  2589--2609.

\bibitem[{Horn and Johnson(2012)}]{horn2012}
Horn, R.~A., Johnson, C.~R., 2012. Matrix Analysis, 2nd Edition. Cambridge
  University Press.

\bibitem[{Howard and Matheson(2005)}]{howard2005}
Howard, R.~A., Matheson, J.~E., 2005. Influence diagrams. Decis. Anal. 2~(3),
  127--143.

\bibitem[{Ibáñez et~al.(2016)Ibáñez, Armañanzas, Bielza, and
  Larrañaga}]{ibanez2015}
Ibáñez, A., Armañanzas, R., Bielza, C., Larrañaga, P., 2016. Genetic
  algorithms and {G}aussian {B}ayesian networks to uncover the predictive core
  set of bibliometric indices. J. Assoc. Inf. Sci. Technol. 67~(7), 1703--1721.

\bibitem[{Isham(1981)}]{isham1981}
Isham, V., 1981. An introduction to spatial point processes and {M}arkov random
  fields. Int. Stat. Rev. 49~(1), 21--43.

\bibitem[{Jones et~al.(2005)Jones, Carvalho, Dobra, Hans, Carter, and
  West}]{jones2005}
Jones, B., Carvalho, C., Dobra, A., Hans, C., Carter, C., West, M., 2005.
  Experiments in stochastic computation for high-dimensional graphical models.
  Stat. Sci. 20~(4), 388--400.

\bibitem[{Jordan(2004)}]{jordan2004}
Jordan, M.~I., 2004. Graphical models. Stat. Sci. 19~(1), 140--155.

\bibitem[{Kalisch and B\"uhlmann(2007)}]{kalisch2007}
Kalisch, M., B\"uhlmann, P., 2007. Estimating high-dimensional directed acyclic
  graphs with the {PC}-algorithm. J. Mach. Learn. Res. 8, 613--636.

\bibitem[{Kiiveri and Speed(1982)}]{kiiveri1982}
Kiiveri, H., Speed, T., 1982. Structural analysis of multivariate data: A
  review. Sociol. Methodol. 13, 209--289.

\bibitem[{Kindermann and Snell(1980)}]{kindermann1980}
Kindermann, R., Snell, J.~L., 1980. Markov Random Fields and their
  Applications. American Mathematical Society.

\bibitem[{Koster(2002)}]{koster2002}
Koster, J.~T., 12 2002. Marginalizing and conditioning in graphical models.
  Bernoulli 8~(6), 817--840.

\bibitem[{Koster(1999)}]{koster1999}
Koster, J. T.~A., 1999. On the validity of the {M}arkov interpretation of path
  diagrams of {G}aussian structural equations systems with correlated errors.
  Scand. J. Stat. 26~(3), 413--431.

\bibitem[{Lam and Fan(2009)}]{lam2009}
Lam, C., Fan, J., 2009. Sparsistency and rates of convergence in large
  covariance matrix estimators. Ann. Stat. 37~(6B), 4254--4278.

\bibitem[{Langseth et~al.(2014)Langseth, Nielsen, P\'erez-Bernab\'e, and
  Salmer\'on}]{langseth2014}
Langseth, H., Nielsen, T., P\'erez-Bernab\'e, I., Salmer\'on, A., 2014.
  Learning mixtures of truncated basis functions from data. Int. J. Approx.
  Reason. 55, 940--956.

\bibitem[{Langseth et~al.(2010)Langseth, Nielsen, Rum\'i, and
  Salmer\'on}]{langseth2010}
Langseth, H., Nielsen, T., Rum\'i, R., Salmer\'on, A., 2010. Parameter
  estimation and model selection for mixtures of truncated exponentials. Int.
  J. Approx. Reason. 51, 485--498.

\bibitem[{Langseth et~al.(2012)Langseth, Nielsen, Rum\'i, and
  Salmer\'on}]{langseth2012}
Langseth, H., Nielsen, T., Rum\'i, R., Salmer\'on, A., 2012. Mixtures of
  truncated basis functions. Int. J. Approx. Reason. 53, 212--227.

\bibitem[{Lauritzen and Sadeghi(2018)}]{lauritzen2018}
Lauritzen, S., Sadeghi, K., 2018. Unifying {{M}arkov} properties for graphical
  models. Ann. Stat. 46~(5), 2251--2278.

\bibitem[{Lauritzen(1992)}]{lauritzen1992}
Lauritzen, S.~L., 1992. Propagation of probabilities, means, and variances in
  mixed graphical association models. J. Amer. Stat. Assoc. 87~(420),
  1098--1108.

\bibitem[{Lauritzen(1996)}]{lauritzen1996}
Lauritzen, S.~L., 1996. Graphical Models. Oxford University Press.

\bibitem[{Lauritzen et~al.(1990)Lauritzen, Dawid, Larsen, and
  Leimer}]{lauritzen1990}
Lauritzen, S.~L., Dawid, A.~P., Larsen, B.~N., Leimer, H.-G., 1990.
  Independence properties of directed {M}arkov fields. Networks 20~(5),
  491--505.

\bibitem[{Lauritzen and Sheehan(2003)}]{lauritzen2003}
Lauritzen, S.~L., Sheehan, N.~A., 11 2003. Graphical models for genetic
  analyses. Stat. Sci. 18~(4), 489--514.

\bibitem[{Lauritzen and Wermuth(1989)}]{lauritzen1989}
Lauritzen, S.~L., Wermuth, N., 1989. Graphical models for associations between
  variables, some of which are qualitative and some quantitative. Ann. Stat.
  17~(1), 31--57.

\bibitem[{Letac and Massam(2007)}]{letac2007}
Letac, G., Massam, H., 2007. {W}ishart distributions for decomposable graphs.
  Ann. Stat. 35~(3), 1278--1323.

\bibitem[{Lin et~al.(2014)Lin, Uhler, Sturmfels, and B\"uhlmann}]{lin2014}
Lin, S., Uhler, C., Sturmfels, B., B\"uhlmann, P., 2014. Hypersurfaces and
  their singularities in partial correlation testing. Found. Comput. Math.
  14~(5), 1079--1116.

\bibitem[{Liu et~al.(2012)Liu, Han, Yuan, Lafferty, and Wasserman}]{liu2012}
Liu, H., Han, F., Yuan, M., Lafferty, J., Wasserman, L., 2012. High-dimensional
  semiparametric {G}aussian copula graphical models. Ann. Stat. 40~(4),
  2293--2326.

\bibitem[{Liu et~al.(2009)Liu, Lafferty, and Wasserman}]{liu2009}
Liu, H., Lafferty, J., Wasserman, L., 2009. The nonparanormal: {S}emiparametric
  estimation of high dimensional undirected graphs. J. Mach. Learn. Res. 10,
  2295--2328.

\bibitem[{Liu(2013)}]{liu2013}
Liu, W., 12 2013. {G}aussian graphical model estimation with false discovery
  rate control. Ann. Stat. 41~(6), 2948--2978.

\bibitem[{Loh and B\"uhlmann(2014)}]{loh2014}
Loh, P.-L., B\"uhlmann, P., 2014. High-dimensional learning of linear causal
  networks via inverse covariance estimation. J. Mach. Learn. Res. 15,
  3065--3105.

\bibitem[{Magwene and Kim(2004)}]{magwene2004}
Magwene, P.~M., Kim, J., Nov 2004. Estimating genomic coexpression networks
  using first-order conditional independence. Genome Biol. 5~(12), R100.

\bibitem[{Markowetz and Spang(2007)}]{markowetz2007}
Markowetz, F., Spang, R., 2007. Inferring cellular networks -- a review. BMC
  Bioinform. 8~(6), S5.

\bibitem[{Massam(2018)}]{massam2018}
Massam, H., 2018. Bayesian inference in graphical {G}aussian models. In:
  Handbook of Graphical Models. CRC Press, pp. 257--282.

\bibitem[{Meek(1995)}]{meek1995}
Meek, C., 1995. Strong completeness and faithfulness in {B}ayesian networks.
  In: Proc. of the Eleventh Conference on Uncertainty in Artificial
  Intelligence. Morgan Kaufmann, San Francisco, pp. 411--418.

\bibitem[{Meinshausen(2008)}]{meinshausen2008}
Meinshausen, N., 2008. A note on the lasso for {G}aussian graphical model
  selection. Stat. Probab. Lett. 78~(7), 880--884.

\bibitem[{Meinshausen and Bühlmann(2006)}]{meinshausen2006}
Meinshausen, N., Bühlmann, P., 2006. High-dimensional graphs and variable
  selection with the lasso. Ann. Stat. 34~(3), 1436--1462.

\bibitem[{Meinshausen and Yu(2009)}]{meinshausen2009}
Meinshausen, N., Yu, B., 2009. Lasso-type recovery of sparse representations
  for high-dimensional data. Ann. Stat. 37~(1), 246--270.

\bibitem[{Moral et~al.(2001)Moral, Rum\'i, and Salmer\'on}]{moral2001}
Moral, S., Rum\'i, R., Salmer\'on, A., 2001. Mixtures of truncated exponentials
  in hybrid {B}ayesian networks. In: Symbolic and Quantitative Approaches to
  Reasoning with Uncertainty. Vol. 2143 of Lecture Notes in Artificial
  Intelligence. Springer, pp. 156--167.

\bibitem[{Mumford and Ramsey(2014)}]{mumford2014}
Mumford, J.~A., Ramsey, J.~D., 2014. {B}ayesian networks for {fMRI}: A primer.
  NeuroImage 86, 573 -- 582.

\bibitem[{Ness et~al.(2016)Ness, Sachs, and Vitek}]{ness2016}
Ness, R.~O., Sachs, K., Vitek, O., 2016. From correlation to causality:
  Statistical approaches to learning regulatory relationships in large-scale
  biomolecular investigations. J. Proteom. Res. 15~(3), 683--690.

\bibitem[{O'Hagan(1995)}]{ohagan1995}
O'Hagan, A., 1995. Fractional {B}ayes factors for model comparison. J. R. Stat.
  Soc. Ser. B Stat. Methodol. 57~(1), 99--138.

\bibitem[{Olkin and Tate(1961)}]{olkin1961}
Olkin, I., Tate, R., 1961. Multivariate correlation models with mixed discrete
  and continuous variables. Ann. Math. Stat. 32~(2), 448--465.

\bibitem[{Pearl(1985)}]{pearl1985b}
Pearl, J., 1985. Bayesian networks: A model of self-activated memory for
  evidential reasoning. Tech. Rep. R-43, University of California, Los Angeles.

\bibitem[{Pearl(1986)}]{pearl1986}
Pearl, J., 1986. Fusion, propagation, and structuring in belief networks.
  Artif. Intell. 29~(3), 241 -- 288.

\bibitem[{Pearl(1988)}]{pearl1988}
Pearl, J., 1988. Probabilistic Reasoning in Intelligent Systems. Morgan
  Kaufmann.

\bibitem[{Pearl(2000)}]{pearl2000}
Pearl, J., 2000. Causaliy: Models, Reasoning and Inference. Cambridge
  University Press.

\bibitem[{Pearl and Paz(1987)}]{pearl1987}
Pearl, J., Paz, A., 1987. Graphoids: A graph-based logic for reasoning about
  relevance relations. In: Advances in Artificial Intelligence. Vol.~2.
  Elsevier, pp. 357--363.

\bibitem[{Peters(2014)}]{peters2014}
Peters, J., 2014. On the intersection property of conditional independence and
  its application to causal discovery. J. Causal Inference 3~(1), 97--108.

\bibitem[{Peters et~al.(2014)Peters, Mooij, Janzing, and
  Sch\"olkopf}]{peters2014b}
Peters, J., Mooij, J.~M., Janzing, D., Sch\"olkopf, B., 2014. Causal discovery
  with continuous additive noise models. J. Mach. Learn. Res. 15, 2009--2053.

\bibitem[{Peña(2016{\natexlab{a}})}]{pena2016}
Peña, J.~M., 2016{\natexlab{a}}. Alternative {M}arkov and causal properties
  for acyclic directed mixed graphs. In: Proc. of the Thirty-Second Conference
  on Uncertainty in Artificial Intelligence. AUAI Press, Arlington, pp.
  577--586.

\bibitem[{Peña(2016{\natexlab{b}})}]{pena2016b}
Peña, J.~M., 2016{\natexlab{b}}. Learning acyclic directed mixed graphs from
  observations and interventions. In: Proc. of the Eighth International
  Conference on Probabilistic Graphical Models. Vol.~52 of Proceedings of
  Machine Learning Research. PMLR, Lugano, pp. 392--402.

\bibitem[{Peña(2018)}]{pena2018}
Peña, J.~M., 2018. Unifying {DAGs} and {UGs}. In: Proc. of the Ninth
  International Conference on Probabilistic Graphical Models. Vol.~72 of
  Proceedings of Machine Learning Research. PMLR, Prague, pp. 308--319.

\bibitem[{Porteous(1989)}]{porteous1989}
Porteous, B.~T., 1989. Stochastic inequalities relating a class of
  log-likelihood ratio statistics to their asymptotic $\chi^2$ distribution.
  Ann. Stat. 17~(4), 1723--1734.

\bibitem[{Radhakrishnan et~al.(2018)Radhakrishnan, Solus, and
  Uhler}]{radhakrishnan2018}
Radhakrishnan, A., Solus, L., Uhler, C., 2018. Counting {M}arkov equivalence
  classes for {DAG} models on trees. Discret. Appl. Math. 244, 170 -- 185.

\bibitem[{Rajaratnam(2012)}]{rajaratnam2012}
Rajaratnam, B., 2012. Comment on: Sequences of regressions and their
  independences. TEST 21~(2), 268--273.

\bibitem[{Rajaratnam et~al.(2008)Rajaratnam, Massam, and
  Carvalho}]{rajaratnam2008}
Rajaratnam, B., Massam, H., Carvalho, C.~M., 12 2008. Flexible covariance
  estimation in graphical {G}aussian models. Ann. Stat. 36~(6), 2818--2849.

\bibitem[{Ravikumar et~al.(2011)Ravikumar, Wainwright, Raskutti, and
  Yu}]{ravikumar2011}
Ravikumar, P., Wainwright, M.~J., Raskutti, G., Yu, B., 2011. High-dimensional
  covariance estimation by minimizing $l_1$-penalized log-determinant
  divergence. Electron. J. Stat. 5, 935--980.

\bibitem[{Richardson(2003)}]{richardson2003}
Richardson, T., 2003. {M}arkov properties for acyclic directed mixed graphs.
  Scand. J. Stat. 30~(1), 145--157.

\bibitem[{Richardson and Spirtes(2002)}]{richardson2002}
Richardson, T., Spirtes, P., 08 2002. Ancestral graph {M}arkov models. Ann.
  Stat. 30~(4), 962--1030.

\bibitem[{Richardson et~al.(2012)Richardson, Robins, and
  Shpitser}]{richardson2012}
Richardson, T.~S., Robins, J.~M., Shpitser, I., 2012. Nested {M}arkov
  properties for acyclic directed mixed graphs. In: Proc. of the Twenty-Eighth
  Conference on Uncertainty in Artificial Intelligence. AUAI Press, Arlington,
  pp. 13--13.

\bibitem[{Robins et~al.(2003)Robins, Scheines, Spirtes, and
  Wasserman}]{robins2003}
Robins, J.~M., Scheines, R., Spirtes, P., Wasserman, L., 2003. Uniform
  consistency in causal inference. Biometrika 90~(3), 491--515.

\bibitem[{Rothman et~al.(2008)Rothman, Bickel, Levina, and Zhu}]{rothman2008}
Rothman, A.~J., Bickel, P.~J., Levina, E., Zhu, J., 2008. Sparse permutation
  invariant covariance estimation. Electron. J. Stat. 2, 494--515.

\bibitem[{Roverato(2000)}]{roverato2000}
Roverato, A., 2000. Cholesky decomposition of a hyper inverse {W}ishart matrix.
  Biometrika 87~(1), 99--112.

\bibitem[{Roverato(2002)}]{roverato2002}
Roverato, A., 2002. Hyper inverse {W}ishart distribution for non-decomposable
  graphs and its application to {B}ayesian inference for {G}aussian graphical
  models. Scand. J. Stat. 29~(3), 391--411.

\bibitem[{Roverato and Whittaker(1998)}]{roverato1998}
Roverato, A., Whittaker, J., 1998. The {I}sserlis matrix and its application to
  non-decomposable graphical {G}aussian models. Biometrika 85~(3), 711--725.

\bibitem[{Sadeghi(2013)}]{sadeghi2013}
Sadeghi, K., 2013. Stable mixed graphs. Bernoulli 19~(5B), 2330--2358.

\bibitem[{Sadeghi(2016)}]{sadeghi2016}
Sadeghi, K., 2016. Marginalization and conditioning for {LWF} chain graphs.
  Ann. Stat. 44~(4), 1792--1816.

\bibitem[{Sadeghi and Lauritzen(2014)}]{sadeghi2014}
Sadeghi, K., Lauritzen, S., 2014. Markov properties for mixed graphs. Bernoulli
  20~(2), 676--696.

\bibitem[{Sadeghi and Marchetti(2012)}]{sadeghi2012b}
Sadeghi, K., Marchetti, G.~M., 2012. {Graphical {M}arkov Models with Mixed
  Graphs in R}. {The {R} J.} 4~(2), 65--73.

\bibitem[{Scutari(2013)}]{scutari2013}
Scutari, M., 2013. On the prior and posterior distributions used in graphical
  modelling. Bayesian Anal. 8~(3), 505--532.

\bibitem[{Shenoy and West(2011)}]{shenoy2011}
Shenoy, P., West, J.~C., 2011. Inference in hybrid {B}ayesian networks using
  mixtures of polynomials. Int. J. Approx. Reason. 52~(5), 641--657.

\bibitem[{Shimizu et~al.(2006)Shimizu, Hoyer, Hyvarinen, and
  Kerminen}]{shimizu2006}
Shimizu, S., Hoyer, P.~O., Hyvarinen, A., Kerminen, A., 2006. A linear
  non-{G}aussian acyclic model for causal discovery. J. Mach. Learn. Res. 7,
  2003--2030.

\bibitem[{Shojaie and Michailidis(2010)}]{shojaie2010}
Shojaie, A., Michailidis, G., 2010. Penalized likelihood methods for estimation
  of sparse high-dimensional directed acyclic graphs. Biometrika 97~(3),
  519--538.

\bibitem[{Sidak(1967)}]{sidak1967}
Sidak, Z., 1967. Rectangular confidence regions for the means of multivariate
  normal distributions. J. Amer. Stat. Assoc. 62~(318), 626--633.

\bibitem[{Sonntag et~al.(2015)Sonntag, Peña, and Gómez-Olmedo}]{sonntag2015}
Sonntag, D., Peña, J.~M., Gómez-Olmedo, M., 2015. Approximate counting of
  graphical models via {MCMC} revisited. Intern. J. Intell. Sys. 30~(3),
  384--420.

\bibitem[{Speed(1979)}]{speed1979}
Speed, T.~P., 1979. A note on nearest-neighbour {G}ibbs and {M}arkov
  probabilities. Sankhya \emph{A} 41~(3/4), 184--197.

\bibitem[{Spirtes(1995)}]{spirtes1995}
Spirtes, P., 1995. Directed cyclic graphical representations of feedback
  models. In: Proc. of the Eleventh Conference on Uncertainty in Artificial
  Intelligence. Morgan Kaufmann, San Francisco, pp. 491--498.

\bibitem[{Spirtes et~al.(2000)Spirtes, Glymour, and Scheines}]{spirtes2000}
Spirtes, P., Glymour, C., Scheines, R., 2000. Causation, Prediction, and
  Search. MIT Press.

\bibitem[{Spirtes et~al.(1997)Spirtes, Richardson, and Meek}]{spirtes1997}
Spirtes, P., Richardson, T., Meek, C., 1997. The dimensionality of mixed
  ancestral graphs. Tech. Rep. CMU-PHIL-83, Carnegie Mellon University.

\bibitem[{Steinsky(2004)}]{steinsky2004}
Steinsky, B., 2004. Asymptotic behaviour of the number of labelled essential
  acyclic digraphs and labelled chain graphs. Graphs Combin. 20~(3), 399--411.

\bibitem[{Studen{\`y}(2005)}]{studeny2005}
Studen{\`y}, M., 2005. On Probabilistic Conditional Independence Structures.
  Springer London.

\bibitem[{Studen{\`y}(2018)}]{studeny2018}
Studen{\`y}, M., 2018. Conditional independence and basic {M}arkov properties.
  In: Handbook of Graphical Models. CRC Press, pp. 21--56.

\bibitem[{Sturmfels and Uhler(2010)}]{sturmfels2010}
Sturmfels, B., Uhler, C., 2010. Multivariate {G}aussians, semidefinite matrix
  completion, and convex algebraic geometry. Ann. Inst. Stat. Math. 62~(4),
  603--638.

\bibitem[{Tibshirani(1996)}]{tibshirani1996}
Tibshirani, R., 1996. Regression shrinkage and selection via the lasso. J. R.
  Stat. Soc. Ser. B Stat. Methodol. 58~(1), 267--288.

\bibitem[{Uhler(2012)}]{uhler2012}
Uhler, C., 2012. Geometry of maximum likelihood estimation in {Gaussian}
  graphical models. Ann. Stat. 40~(1), 238--261.

\bibitem[{Uhler(2018)}]{uhler2018b}
Uhler, C., 2018. Gaussian graphical models. In: Handbook of Graphical Models.
  CRC Press, pp. 235--256.

\bibitem[{Uhler et~al.(2018)Uhler, Lenkoski, and Richards}]{uhler2018}
Uhler, C., Lenkoski, A., Richards, D., 2018. Exact formulas for the normalizing
  constants of {W}ishart distributions for graphical models. Ann. Stat. 46~(1),
  90--118.

\bibitem[{Uhler et~al.(2013)Uhler, Raskutti, B\"uhlmann, and Yu}]{uhler2013}
Uhler, C., Raskutti, G., B\"uhlmann, P., Yu, B., 2013. Geometry of the
  faithfulness assumption in causal inference. Ann. Stat. 41~(2), 436--463.

\bibitem[{van~de Geer and Bühlmann(2013)}]{geer2013}
van~de Geer, S., Bühlmann, P., 2013. $\ell_{0}$-penalized maximum likelihood
  for sparse directed acyclic graphs. Ann. Stat. 41~(2), 536--567.

\bibitem[{van~de Geer and B\"uhlmann(2009)}]{geer2009}
van~de Geer, S.~A., B\"uhlmann, P., 2009. On the conditions used to prove
  oracle results for the lasso. Electron. J. Stat. 3, 1360--1392.

\bibitem[{Varando et~al.(2015)Varando, L\'opez-Cruz, Nielsen, Larrañaga, and
  Bielza}]{varando2015}
Varando, G., L\'opez-Cruz, P., Nielsen, T., Larrañaga, P., Bielza, C., 2015.
  Conditional density approximations with mixtures of polynomials. Intern. J.
  Intell. Sys. 30, 236--264.

\bibitem[{Verma and Pearl(1991)}]{verma1991}
Verma, T., Pearl, J., 1991. Equivalence and synthesis of causal models. In:
  Proc. of the Sixth Annual Conference on Uncertainty in Artificial
  Intelligence. AUAI Press, Corvallis, pp. 255--270.

\bibitem[{Vogel and Fried(2011)}]{vogel2011}
Vogel, D., Fried, R., 2011. Elliptical graphical modelling. Biometrika 98~(4),
  935--951.

\bibitem[{Vogel and Tyler(2014)}]{vogel2014}
Vogel, D., Tyler, D.~E., 2014. Robust estimators for nondecomposable elliptical
  graphical models. Biometrika 101~(4), 865.

\bibitem[{Wang and Carvalho(2010)}]{wang2010}
Wang, H., Carvalho, C.~M., 2010. Simulation of hyper-inverse {W}ishart
  distributions for non-decomposable graphs. Electron. J. Stat. 4, 1470--1475.

\bibitem[{Werhli et~al.(2006)Werhli, Grzegorczyk, and Husmeier}]{werhli2006}
Werhli, A.~V., Grzegorczyk, M., Husmeier, D., 2006. Comparative evaluation of
  reverse engineering gene regulatory networks with relevance networks,
  graphical {G}aussian models and {B}ayesian networks. Bioinformatics 22~(20),
  2523--2531.

\bibitem[{Wermuth(1976{\natexlab{a}})}]{wermuth1976}
Wermuth, N., 1976{\natexlab{a}}. Analogies between multiplicative models in
  contingency tables and covariance selection. Biometrics 32~(1), 95--108.

\bibitem[{Wermuth(1976{\natexlab{b}})}]{wermuth1976b}
Wermuth, N., 1976{\natexlab{b}}. Model search among multiplicative models.
  Biometrics 32~(2), 253--263.

\bibitem[{Wermuth(1980)}]{wermuth1980}
Wermuth, N., 1980. Linear recursive equations, covariance selection, and path
  analysis. J. Am. Stat. Assoc. 75~(372), 963--972.

\bibitem[{Wermuth(2011)}]{wermuth2011}
Wermuth, N., 2011. Probability distributions with summary graph structure.
  Bernoulli 17~(3), 845--879.

\bibitem[{Wermuth(2015)}]{wermuth2015}
Wermuth, N., 2015. Graphical {M}arkov models, unifying results and their
  interpretation. Wiley StatsRef: Stat. Ref. Online, 1--29.

\bibitem[{Wermuth and Lauritzen(1983)}]{wermuth1983}
Wermuth, N., Lauritzen, S.~L., 1983. Graphical and recursive models for
  contingency tables. Biometrika 70~(3), 537--552.

\bibitem[{Wermuth and Sadeghi(2012)}]{wermuth2012}
Wermuth, N., Sadeghi, K., 2012. Sequences of regressions and their
  independences. TEST 21~(2), 215--252.

\bibitem[{Wille and Bühlmann(2006)}]{wille2006}
Wille, A., Bühlmann, P., 2006. Low-order conditional independence graphs for
  inferring genetic networks. Stat. Appl. Genet. Mol. Biol. 5~(1), {A}rticle 1.

\bibitem[{Wright(1934)}]{wright1934}
Wright, S., 1934. The method of path coefficients. Ann. Math. Stat. 5~(3),
  161--215.

\bibitem[{Xue and Zou(2012)}]{xue2012}
Xue, L., Zou, H., 2012. Regularized rank-based estimation of high-dimensional
  nonparanormal graphical models. Ann. Stat. 40~(5), 2541--2571.

\bibitem[{Yu and Bien(2017)}]{yu2017}
Yu, G., Bien, J., 2017. Learning local dependence in ordered data. J. Mach.
  Learn. Res. 18, 1--60.

\bibitem[{Yuan and Lin(2007{\natexlab{a}})}]{yuan2007}
Yuan, M., Lin, Y., 2007{\natexlab{a}}. Model selection and estimation in the
  {Gaussian} graphical model. Biometrika 94~(1), 19--35.

\bibitem[{Yuan and Lin(2007{\natexlab{b}})}]{yuan2007b}
Yuan, M., Lin, Y., 2007{\natexlab{b}}. On the non-negative garrotte estimator.
  J. R. Stat. Soc. Ser. B Stat. Methodol. 69~(2), 143--161.

\bibitem[{Yule(1907)}]{yule1907}
Yule, G.~U., 1907. On the theory of correlation for any number of variables,
  treated by a new system of notation. Proc. R. Soc. \emph{A} 79~(529),
  182--193.

\bibitem[{Zhang and Spirtes(2003)}]{zhang2003}
Zhang, J., Spirtes, P., 2003. Strong faithfulness and uniform consistency in
  causal inference. In: Proc. of the Nineteenth Conference on Uncertainty in
  Artificial Intelligence. Morgan Kaufmann, San Francisco, pp. 632--639.

\bibitem[{Zhao and Yu(2006)}]{zhao2006}
Zhao, P., Yu, B., 2006. On model selection consistency of lasso. J. Mach.
  Learn. Res. 7, 2541--2563.

\bibitem[{Zhou et~al.(2016)Zhou, Wang, Liu, Ogunbona, and Dinggang}]{zhou2016}
Zhou, L., Wang, L., Liu, L., Ogunbona, P., Dinggang, S., 2016. Learning
  discriminative {B}ayesian networks from high-dimensional continuous
  neuroimaging data. IEEE Trans. Pattern Anal. Mach. Intell. 38~(11),
  2269--2283.

\bibitem[{Zou(2006)}]{zou2006}
Zou, H., 2006. The adaptive lasso and its oracle properties. J. Amer. Stat.
  Assoc. 101~(476), 1418--1429.

\end{thebibliography}

\end{document}